\documentclass[aoas]{imsart}

\RequirePackage{amsthm,amsmath,amsfonts,amssymb}
\RequirePackage[authoryear]{natbib}
\RequirePackage[colorlinks,citecolor=blue,urlcolor=blue]{hyperref}
\RequirePackage{graphicx}
\usepackage{multicol,multirow}
\usepackage{subfig, caption, float}
\usepackage{bbm}

\makeatletter
\newcommand*\bigcdot{\mathpalette\bigcdot@{.5}}
\newcommand*\bigcdot@[2]{\mathbin{\vcenter{\hbox{\scalebox{#2}{$\m@th#1\bullet$}}}}}
\makeatother

\usepackage[ruled]{algorithm2e}
\SetKwInput{KwParam}{Parameter}
\SetAlgoCaptionLayout{centerline}

\startlocaldefs
\newcommand{\indep}{\rotatebox[origin=c]{90}{$\models$}}

\newtheorem{theorem}{Theorem}[section]

\newtheorem{remark}{Remark}
\newtheorem{assumption}{Assumption}
\theoremstyle{remark}
\newtheorem{definition}[theorem]{Definition}
\newtheorem*{example}{Example}

\endlocaldefs

\begin{document}

\begin{frontmatter}
\title{Social Distancing and COVID-19: Randomization Inference for a Structured Dose-Response Relationship}
\runtitle{Social Distancing and COVID-19}
\begin{aug}
\author[A]{\fnms{Bo} \snm{Zhang}\ead[label=e1]{bozhan@wharton.upenn.edu}},
\author[B]{\fnms{Siyu} \snm{Heng}\ead[label=e2]{siyuheng@sas.upenn.edu}},
\author[A]{\fnms{Ting} \snm{Ye}\ead[label=e3,mark]{tingye@wharton.upenn.edu}},
\and
\author[A]{\fnms{Dylan S.} \snm{Small}\ead[label=e4,mark]{dsmall@wharton.upenn.edu}}
\address[A]{Department of Statistics, The Wharton School, University of Pennsylvania,
\printead{e1,e3,e4}}

\address[B]{Graduate Group in Applied Mathematics and Computational Science, School of Arts and Sciences, University of Pennsylvania,
\printead{e2}}
\end{aug}

\begin{abstract}
Social distancing is widely acknowledged as an effective public health policy combating the novel coronavirus. But extreme forms of social distancing like isolation and quarantine have costs and it is not clear how much social distancing is needed to achieve public health effects. In this article, we develop a design-based framework to test the causal null hypothesis and make inference about the dose-response relationship between reduction in social mobility and COVID-19 related public health outcomes. We first discuss how to embed observational data with a time-independent, continuous treatment dose into an approximate randomized experiment, and develop a randomization-based procedure that tests if a structured dose-response relationship fits the data. We then generalize the design and testing procedure to accommodate a time-dependent treatment dose in a longitudinal setting. Finally, we apply the proposed design and testing procedures to investigate the effect of social distancing during the phased reopening in the United States on public health outcomes using data compiled from sources including $\textsf{Unacast}\textsuperscript{\texttrademark}$, the United States Census Bureau, and the County Health Rankings and Roadmaps Program. We rejected a primary analysis null hypothesis that stated the social distancing from April 27, 2020, to June 28, 2020, had no effect on the COVID-19-related death toll from June 29, 2020, to August 2, 2020 (p-value $< 0.001$), and found that it took more reduction in mobility to prevent exponential growth in case numbers for non-rural counties compared to rural counties.
\end{abstract}

\begin{keyword}
\kwd{causal inference}
\kwd{COVID-19}
\kwd{longitudinal studies}
\kwd{randomization inference}
\kwd{statistical matching}
\kwd{dose-response}
\end{keyword}

\end{frontmatter}

\section{Introduction}
\label{sec: introduction}
\subsection{Social distancing, a pilot study, and dose-response relationship}
Social distancing is widely acknowledged as one of the most effective public health strategies to reduce transmission of the novel coronavirus (\citealp{lewnard2020scientific}). There seemed to be ample evidence from China (\citealp{lau2020positive}) and Italy (\citealp{sjodin2020only}) that a strict lockdown and practice of social distancing could have a substantial effect on reducing disease transmission, but social distancing has economic, psychological and societal costs (\citealp{acemoglu2020multi,atalan2020lockdown, grover2020psychological,sheridan2020social,  venkatesh2020social}). How much social distancing is needed to achieve the desired public health effect? In this article, we measure the level of social distancing using data on daily percentage change in total distance traveled compared to the pre-coronavirus level (data compiled and made available by $\textsf{Unacast}\textsuperscript{\texttrademark}$) and investigate the causal relationship between social distancing and COVID-related public health outcomes.

We conducted a pilot study in March to investigate the effect of social distancing during the first week of President Trump's 15 Days to Slow the Spread campaign (March 16-22, 2020) on the influenza-like illness (ILI) percentage two and three weeks later. We tested the causal null hypothesis and found some weak evidence (p-value $=0.08$) that better social distancing had an effect on ILI percentage three weeks later. In Supplementary Material A, we described in detail our pilot study. A protocol of the design and analysis was posted on \textsf{arXiv} (\url{https://arxiv.org/abs/2004.02944}) before outcome data were available and analyzed.

In addition to the causal null hypothesis, the ``dose-response relationship" between the degree of social distancing and potential public health outcomes under various degrees of social distancing is also of great interest. Infectious disease experts seemed to express sentiments that the effect of social distancing on public health outcomes might be small or even negligible under a small degree of social distancing, but much more substantial under a large degree of social distancing. In an interview with the British Broadcasting Corporation (\citealp{Fauci2020interview}), director of the National Institute of Allergy and Infectious Diseases (NIAID), Dr. Anthony S. Fauci said:
\begin{quote}
    ``We never got things down to baseline where so many countries in Europe and the UK and other countries did {\textendash} they closed down to the tune of about \textbf{97 percent} lockdown. In the United States, even in the most strict lockdown, only about \textbf{50 percent} of the country was locked down. That allowed the perpetuation of the outbreak that we never did get under very good control".
\end{quote}
Perhaps Dr. Fauci was proposing a \emph{hypothesis} that the treatment dose, i.e., level of social distancing, played a very important role, and the causal effect of social distancing as a public health strategy combating coronavirus transmission is likely to be very different depending on the extent to which it is practiced (see, e.g., \citet{gelfand2021relationship}). We would like to formalize and test the hypothesis concerning a dose-response relationship between social distancing and public health outcomes.

\subsection{Reopening, causal null hypothesis, dose-response kink model, and connection to epidemiological models}
\begin{figure}[h]
    \centering
    \includegraphics[width = 0.78\columnwidth]{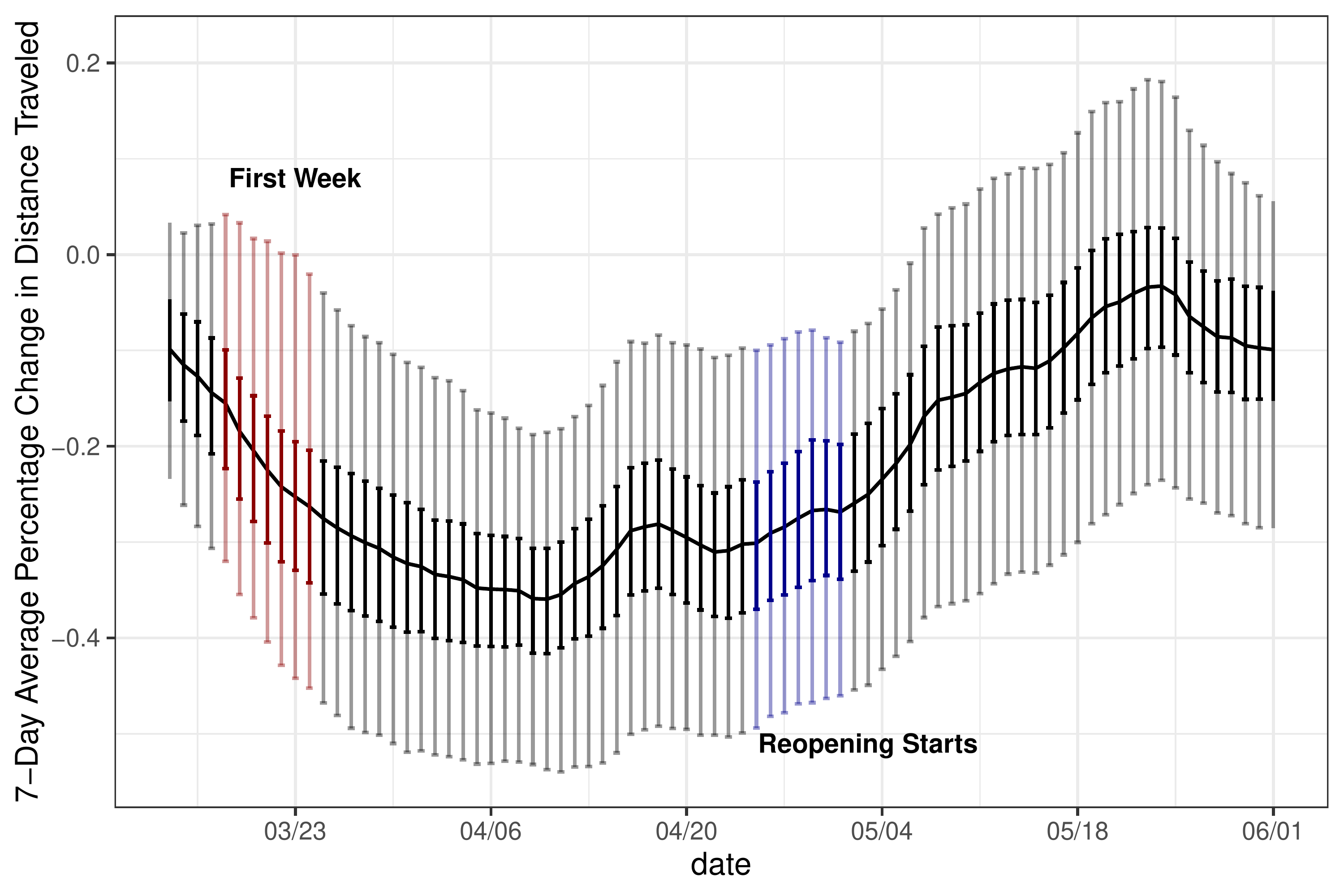}
    \caption{\small County-averaged 7-day rolling average (black solid line), middle $50\%$ (dark shade), and middle $90\%$ (light shade) of percentage change in total distance traveled, e.g., $-0.35$ corresponds to $35\%$ reduction in total distance traveled compared to the pre-coronavirus period. The first week of 15 Days to Slow the Spread campaign (March 16-22) is marked in red and the first week of reopening in blue.}
    \label{fig: social distancing vs time}
\end{figure}

Starting late April and early May, many states in the U.S. started phased reopening. States and local governments differed in their reopening timelines; people in different states and counties also differed in their social mobility during the process: some ventured out; some continued to stay at home as much as possible. Figure \ref{fig: social distancing vs time} plots the 7-day rolling average of percentage change in total distance traveled of all counties in the U.S., from mid-March to late May. It is evident that as many counties started to ease social distancing measures, we saw less reduction in distance traveled; in fact, in many counties, distance traveled started to return to and even supersede the pre-coronavirus level. 

\begin{figure}[h]
    \centering
    \includegraphics[width=0.8\textwidth]{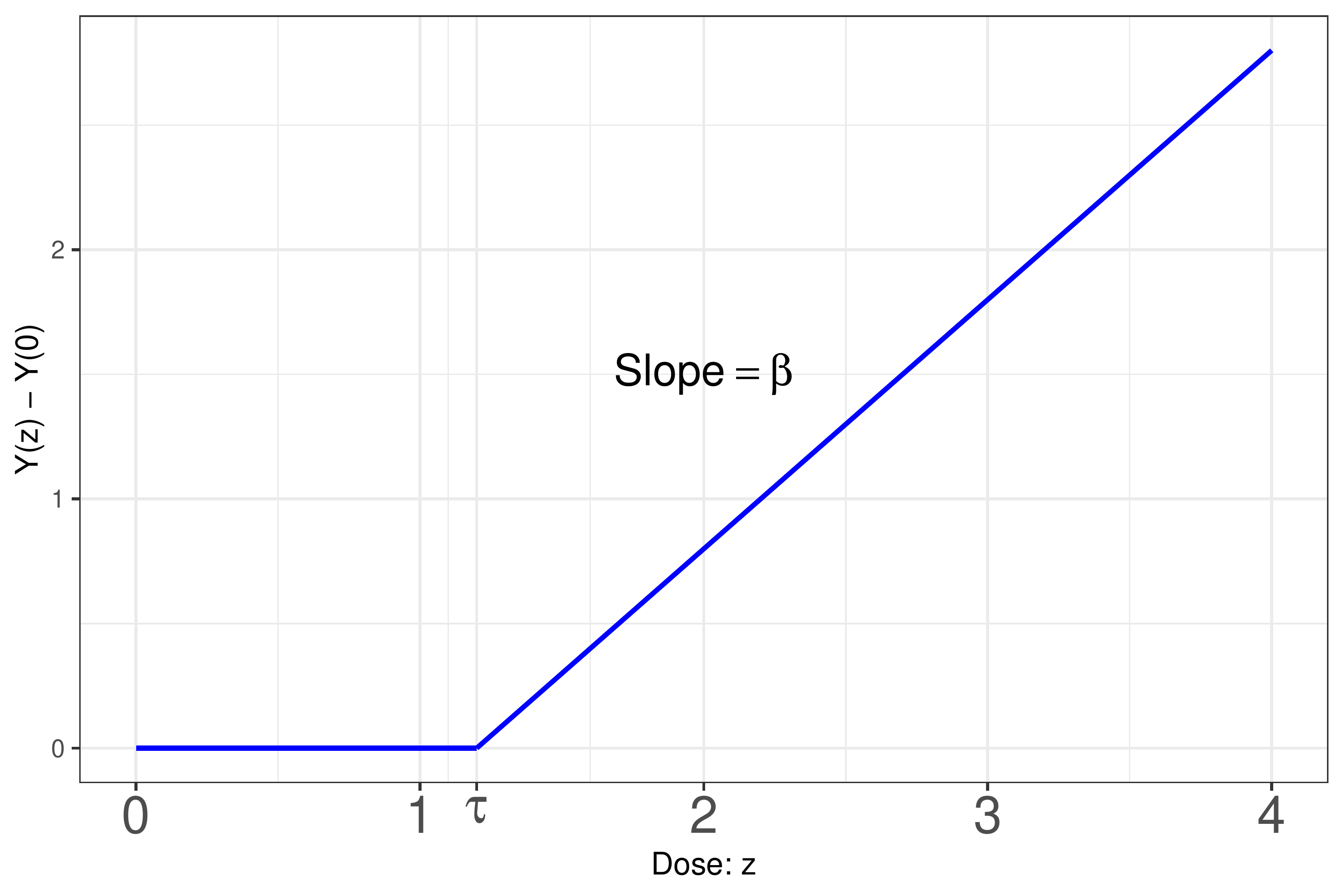}
    \caption{The dose-response kink model}
    \label{fig: kink model}
\end{figure}

In this article, we leverage the county-level social mobility data since phased reopening in the U.S. to study the relationship between social mobility and its effect on public health outcomes. Let $t_0$ denote a baseline period, $T$ some endpoint of interest, $\mathbf{z}_{t_0:T}$ a longitudinal measurement of change in social mobility in county $n$ from $t_0$ to $T$, and $Y_{n, T}(\mathbf{z}_{t_0:T})$ county $n$'s potential public health outcome at time $T$ under the social mobility trajectory $\mathbf{z}_{t_0:T}$, e.g., the number of patients succumbing to the COVID-19 at time $T$. Our first scientific query is about the causal null hypothesis: had the social mobility trajectory changed from $\mathbf{z}_{t_0: T}$ to $\mathbf{z}'_{t_0: T}$, would the potential public health outcome at time $T$ change at all? In other words, does $Y_{n, T}(\mathbf{z}_{t_0: T}) = Y_{n, T}(\mathbf{z}'_{t_0: T})$ hold for all $\mathbf{z}_{t_0:T} \neq \mathbf{z}'_{t_0:T}$? Suppose that we have enough evidence from observational data to reject this causal null hypothesis, our second query then is about the dose-response relationship between the level of social distancing and its effect on the potential public health outcome. To illustrate, one such dose-response relationship (among many other candidates) is the following dose-response kink model (see Figure \ref{fig: kink model} for an illustration):
\begin{equation}
\label{eqn: dose-response kink model}
\begin{split}
H_0^K:~\quad
    &Y_{n, T}(z) = Y_{n, T}(z^\ast),~\forall z \leq \tau,~\text{and}\\
    &Y_{n, T}(z) - Y_{n, T}(\tau) = \beta(z - \tau),~\forall z > \tau, ~\forall n = 1, 2, \cdots, N,\\
    &\text{for some } \tau ~\text{and}~\beta,
\end{split}
\end{equation}
where $z$ captures some aggregate dose of the social mobility trajectory $\mathbf{z}_{t_0:T}$, e.g., the average reduction in social mobility from $t_0$ to $T$, and $z^\ast$ a reference dose level. Model \eqref{eqn: dose-response kink model} states that the potential health-related outcome (e.g., daily death toll, test positivity rate, etc) at time $T$ would remain unchanged as the potential outcome under the reference level when the aggregate dose $z$ is less than a certain threshold $\tau$, and then increases at a rate proportional to how much $z$ exceeds the threshold. Model \eqref{eqn: dose-response kink model} succinctly captures two key features policy makers may be most interested in: $\tau$ the minimum dose that ``activates" the treatment effect, and $\beta$ how fast the potential outcome changes as the dose changes after exceeding the threshold. Model \eqref{eqn: dose-response kink model} may remind readers of the ``broken line regression" models in regression analysis; see \citet[Chapter 4]{zhang2010recursive}. The key difference here is that Model \eqref{eqn: dose-response kink model} and other dose-response relationships in this article are about the contrast in potential outcomes, not the observed outcomes in a regression analysis. 

Our analysis in this article complements standard analyses based on epidemiological models, e.g., the SIR (susceptible-infected-recovered) compartment models (\citealp{brauer2012mathematical}). The primary interest of epidemiological models is to understand infectious disease dynamics, in particular how the public health outcome trajectory evolves over time. To investigate a dose-response relationship, we only posit a parsimonious model on the contrast between potential outcomes at time $T$ under different doses, e.g., $Y_{n, T}(z) - Y_{n, T}(\tau)$ in \eqref{eqn: dose-response kink model}, not on the disease dynamics that generate the outcome $Y_{n, T}(z)$. In other words, a parsimonious dose-response relationship does not preclude nonlinear infectious disease dynamics, e.g., those based on the compartment models; moreover, our primary inferential target, the causal null hypothesis, does not impose any restriction on the infectious disease mechanism. 

\subsection{Our contribution}
We have three goals in this article. First, we propose a simple, model-free randomization-based procedure that tests if a causal null hypothesis or a structured dose-response relationship, e.g., the dose-response kink model, fits the data in a static setting with a time-independent, continuous or many-leveled treatment dose. To be specific, an empirical researcher posits a structured dose-response relationship that she finds scientifically meaningful, parsimonious, and flexible enough to describe data at hand; our developed procedure can then be applied to test if such a postulated dose-response relationship is sufficient to describe the causal relationship. If the hypothesis is rejected, empirical researchers are then advised to re-examine the scientific theory underpinning the postulated model; otherwise, the model seems a good starting point for data analysis. In this way, our method can be deemed as a model-free ``diagnostic test" for a dose-response relationship, and more broadly a test of the underlying scientific theory. In our application, the treatment and outcome are both longitudinal. Our second goal is to generalize the proposed design and testing procedure to the longitudinal setting. We define a notion of cumulative dose for a time-varying treatment dose trajectory, and discuss how to embed observational longitudinal data into an approximate randomized controlled trial in order to permute two treatment trajectories. Finally, we closely examine our assumptions in the context of an infectious disease transmission mechanism and apply the developed design and testing procedure to characterize the dose-response relationship between reduction in social mobility and public health outcomes during the reopening phases in the U.S. using county-level data we compiled from sources including $\textsf{Unacast}\textsuperscript{\texttrademark}$, the United States Census Bureau, and the County Health Rankings and Roadmaps Program (\citealp{remington2015county}). 

The rest of the article is organized as follows. Section 2 and 3 study how to investigate a dose-response relationship using nonbipartite matching in a static setting. Section 4 incorporates interference and considers the spillover effects. Section 5 extends the method to longitudinal studies. Section 6 describes the design of the case study and Section 7 presents results and extensive sensitivity analyses. Section 8 concludes with a discussion.

\section{Investigating the dose-response relationship via nonbipartite matching}
\label{sec: setup nbpmatching and potential outcome}
\subsection{Observational data with a continuous treatment dose in a static setting}
\label{subsec: observational data with continuous treatment}
Suppose there are $N = 2I$ units, indexed by $n = 1, 2, \cdots, N$. Each unit is associated with a vector of observed covariates $\mathbf{X}_n$, an observed treatment dose assignment $Z^{\text{obs}}_n$, and an observed outcome $Y^{\text{obs}}_n$. The vector of observed covariates $\mathbf{X}_n$ are collected before the treatment assignment and not affected by the treatment. Let $Z_n$ denote the treatment dose assignment of unit $n$, $\mathcal{Z}$ the set of all possible treatment doses, $z \in \mathcal{Z}$ a realization of $Z_n$, and $|\mathcal{Z}|$ the cardinality of $\mathcal{Z}$. For a binary treatment, $|\mathcal{Z}| = 2$; for a continuous treatment dose, $|\mathcal{Z}|$ is an infinite number. In most applications, $\mathcal{Z}$ is an ordered set (either partially ordered or totally ordered) with a (partial or total) order defined in light of the application. 

Let $Y_n(z)$ denote the potential outcome that unit $n$ exhibits under the dose assignment $z$ assuming no interference among units (\citealp{rubin1980randomization, rubin1986statistics}). Each unit $n$ is associated with a possibly infinite array of potential outcomes $\{Y_{n}(z),~z\in\mathcal{Z}\}$. We will assume consistency so that $Y_n^{\text{obs}} = Y_n(Z^{\text{obs}}_n)$. A \emph{causal estimand} is necessarily a contrast between potential outcomes. Each unit $n$ is associated with a collection of unit-level causal effects $\{f_{n}(z, z') = Y_{n}(z) - Y_{n}(z'), ~\forall z, z' \in \mathcal{Z}\}$. Table \ref{tbl: science table before matching} summarizes all information regarding these $N$ units, where we let $\mathcal{Z} = \{0, 1, 2, \cdots\}$ be a countable set for ease of exposition. Table \ref{tbl: science table before matching} is referred to as a \emph{science table} in the literature (\citealp{rubin2005causal}). In a causal inference problem, the fundamental estimands of interest are the arrays of potential outcomes in Table \ref{tbl: science table before matching}; the task of uncovering the arrays of potential outcomes is challenging because one and only one of the potentially infinite array of potential outcomes for each unit is actually observed.

\begin{table}[ht]
\centering
\caption{\small Science table of $N = 2I$ units for a countable set $\mathcal{Z} = \{0, 1, 2, \cdots\}$.}
\label{tbl: science table before matching}
\resizebox{\textwidth}{!}{
\fbox{%
\begin{tabular}{ccccccccccc} \\[-0.8em]
 & & &\multicolumn{5}{c}{Potential Outcomes} \\ \\[-0.9em]  \cline{4-8}
\multirow{2}{*}{\begin{tabular}{c}Units\end{tabular}}  & \multirow{2}{*}{\begin{tabular}{c}Covariates\\ $\mathbf{X}$\end{tabular}} & \multirow{2}{*}{\begin{tabular}{c}Observed\\Dose $Z$\end{tabular}} & \multirow{2}{*}{\begin{tabular}{c}$Y(0)$\end{tabular}} & \multirow{2}{*}{\begin{tabular}{c}$Y(1)$\end{tabular}} & \multirow{2}{*}{\begin{tabular}{c}$\cdots$\end{tabular}} & \multirow{2}{*}{\begin{tabular}{c}$Y(z')$\end{tabular}} & \multirow{2}{*}{\begin{tabular}{c}$\cdots$\end{tabular}} &\multirow{2}{*}{\begin{tabular}{c}Unit-Level \\ Causal Effects\end{tabular}}   
&\multirow{2}{*}{\begin{tabular}{c}Unit-level Causal \\ Effects Summary\end{tabular}} &\multirow{2}{*}{\begin{tabular}{c}Summary \\ Causal Effects\end{tabular}}\\ \\[-0.9em]  \\ \hline \\[-0.9em]
$1$ & $\mathbf{X}_1$ & $Z^{\text{obs}}_1$ & $Y_{1}(0)$ & $Y_{1}(1)$  & $\cdots$ & $Y_{1}(z')$ & $\cdots$ & $\{Y_{1}(z) - Y_{1}(z'),~z,z'\in\mathcal{Z}\}$ & \multirow{6}{*}{\begin{tabular}{c}Unit-Level \\ Dose-response \\ relationship\\e.g.,\\$Y_n(z) - Y_n(z^\ast)$ \\
$= f_n(z; z^\ast,\mathbf{\theta_n})$ \end{tabular}}
& \multirow{6}{*}{\begin{tabular}{c}Summarize \\dose-response \\ relationship for \\ a common set \\ of units \end{tabular}}
\\
$2$ & $\mathbf{X}_2$ & $Z^{\text{obs}}_2$ & $Y_{2}(0)$ & $Y_{2}(1)$  & $\cdots$ & $Y_{2}(z')$ & $\cdots$ & $\{Y_{2}(z) - Y_{2}(z'),~z,z'\in\mathcal{Z}\}$\\
$\vdots$ & $\vdots$ &$\vdots$ & $\vdots$ & $\vdots$ & $\vdots$ & $\vdots$ & $\vdots$ & $\vdots$\\
$n$ & $\mathbf{X}_n$ & $Z^{\text{obs}}_n$ & $Y_{n}(0)$ & $Y_{n}(1)$  & $\cdots$ & $Y_{n}(z')$ & $\cdots$ & $\{Y_{n}(z) - Y_{n}(z'),~z,z'\in\mathcal{Z}\}$\\
$\vdots$ & $\vdots$ &$\vdots$ & $\vdots$ & $\vdots$ & $\vdots$ & $\vdots$ & $\vdots$ &$\vdots$\\
$N$ & $\mathbf{X}_N$ &$Z^{\text{obs}}_N$ & $Y_{N}(0)$ & $Y_{N}(1)$  & $\cdots$ & $Y_{N}(z')$ & $\cdots$ & $\{Y_{N}(z) - Y_{N}(z'),~z,z'\in\mathcal{Z}\}$\\
\end{tabular}}}
\end{table}

One unique feature of problems with a continuous treatment dose assignment is that the unit-level causal effect is an infinite set of comparisons between any two potential outcomes $Y_n(z)$ and $Y_n(z')$, unlike with a binary treatment where the unit-level causal effect unambiguously refers to a comparison between $Y_n(1)$ and $Y_n(0)$. Let $z^\ast\in \mathcal{Z}$ denote an arbitrary reference dose. Observe that $Y_n(z) - Y_n(z') = Y_n(z) - Y_n(z^\ast) - \{Y_n(z') - Y_n(z^\ast)\}$, and the collection of contrasts $\{Y_n(z) - Y_n(z^\ast), z\in\mathcal{Z}\}$ is sufficient in summarizing all pairwise comparisons of potential outcomes. With a binary treatment, a ``summary causal effect" (\citealp{rubin2005causal}) is defined as a comparison between $Y_n(1)$ and $Y_n(0)$ over the same collection of units, e.g., the mean unit-level difference for females. With a continuous treatment dose, we first summarize the causal effects with a ``unit-level dose-response relationship" for each unit $n$. For example, one simple unit-level dose-response relationship states that $Y_n(z) - Y_n(z^\ast) = \tau_0,~\forall z\in\mathcal{Z}$; in words, for unit $n$, the causal effect when comparing treatment dose $z$ to the reference dose $z^\ast$ is equal to a constant $\tau_0$ regardless of the dose $z \in \mathcal{Z}$. We may then summarize such unit-level dose-response relationships for a collection of units. For example, one such summary may state that a structured dose-response relationship $f(z; z^\ast,\mathbf{\theta})$ holds for all counties in the U.S.; this summary can be represented by the following null hypothesis:
\begin{equation*}
    H_0^1: Y_n(z) - Y_n(z^\ast) = f(z; z^\ast, \mathbf{\theta}),~\text{for all counties in the U.S. indexed by}~n,~\text{for some}~\mathbf{\theta}.
\end{equation*}
We first develop a simple, randomization-based testing procedure to assess hypotheses of the form $H_0^1$. The work most relevant to our development is \citet{Ding2016variation}, who studied testing the existence of treatment effect variation in a randomized controlled trial with a time-independent binary treatment.

In a randomization-based inferential procedure, the potential outcomes (i.e., the infinite collection $\{Y_n(z),~\forall z\in\mathcal{Z},\forall n\}$ in Table \ref{tbl: science table before matching}, are held fixed and the only probability distribution that enters statistical inference is the randomization distribution that describes the treatment dose assignment. The key step here is to properly embed the observational data into an approximately randomized experiment (\citealp{rosenbaum2002observational,rosenbaum2010design, bind2019bridging}), as we are ready to describe.

\subsection{Embedding observational data with a time-independent, continuous treatment into an as-if randomized experiment via nonbipartite matching}
\label{subsec: embed observational data into experiment}
In a randomized controlled experiment, physical randomization creates ``the reasoned basis" for drawing causal inference (\citealp{Fisher1935design}). In the absence of physical randomization as with retrospective observational data, one strategy is to use statistical matching to embed observational data into a hypothetical randomized controlled trial (\citealp{rosenbaum2002observational,rosenbaum2010design, rubin2007design,ho2007matching, stuart2010matching, bind2019bridging}) by matching subjects with the same (or at least very similar) estimated propensity score or observed covariates and forging two groups that are well-balanced in observed covariates.

One straightforward design to handle observational data with a continuous treatment is to dichotomize the continuous treatment based on some prespecified threshold and create a binary treatment out of the dichotomization scheme. For instance, let $Z$ denote a measure of social distancing; one can define counties with the social distancing measure above the median as the ``above-median," or ``treated" group, and the others as the ``below-median," or ``control" (or ``comparison") group. One may then pair counties in the ``above-median" group to those in the ``below-median" group via a standard bipartite matching algorithm (for instance, via the \textsf{R} package \textsf{optmatch} by \citealp{hansen2007optmatch}), and test the null hypothesis that social distancing has no effect on the outcome. Such a strategy is often seen in empirical research, probably because of its simplicity; however, dichotomizing the continuous treatment inevitably censors the rich information contained in the original, continuous dose and prevents researchers from studying the dose-response relationship.

To address this limitation, \citet{lu2001matching,lu2011optimal} proposed optimal nonbipartite matching. In a nonbipartite matching, units with similar observed covariates but different treatment doses are paired. Suppose there are $N = 2I$ units, e.g., counties in the U.S in our application. In the design stage, distances $\{\delta_{ij},~i = 1, \cdots, N,~j = 1, \cdots, N\}$ are calculated between each pair of units and a $N \times N$ distance matrix is constructed (\citealp{lu2001matching, lu2011optimal,baiocchi2010building}). Some commonly used distances $\delta_{ij}$ include the Mahalanobis distance between observed covariates $\mathbf{X}_i$ and $\mathbf{X}_j$ and the rank-based robust Mahalanobis distance. Researchers may further modify the distance to incorporate specific design aspects of the study. For instance, in a study that involves effect modification, researchers are advised to match exactly or near-exactly on the effect modifier (\citealp{rosenbaum2005heterogeneity}), e.g., the geographic location of the county, and such an aspect of design can be pursued by adding a large penalty to $\delta_{ij}$ if county $i$ and $j$ are not from the same geographic region. 

An optimal nonbipartite matching algorithm then divides these $N = 2I$ units into $I$ non-overlapping pairs of two units such that the total within-matched-pair distance is minimized. Nonbipartite matching allows more flexible pairing compared to bipartite matching based on a dichotomization scheme, and preserves the continuous nature of the treatment, which is essential for investigating a dose-response relationship. 

Suppose that we have formed $I$ matched pairs of $2$ units so that index $ij,~i = 1, \cdots, I,~j = 1, 2,$ uniquely identifies a unit. We follow \citet{rosenbaum1989sensitivity} and \citet{heng2019instrumental} and define the following potential outcomes after nonbipartite matching.

\begin{definition}[Potential Outcomes After Nonbipartite Matching]\rm
\label{def: define potential outcomes}
Let $Z^{\text{obs}}_{i1} \vee Z^{\text{obs}}_{i2} = \max(Z^{\text{obs}}_{i1}, Z^{\text{obs}}_{i2})$ and $Z^{\text{obs}}_{i1} \wedge Z^{\text{obs}}_{i2} = \min(Z^{\text{obs}}_{i1}, Z^{\text{obs}}_{i2})$ denote the maximum and minimum of two observed treatment doses in each matched pair $i$. We define the following two potential outcomes for each unit $ij$:
\begin{equation*}
\begin{split}
    &Y_{Tij} \overset{\Delta}{=} Y_{ij}(Z^{\text{obs}}_{i1} \vee Z^{\text{obs}}_{i2}), \qquad Y_{Cij} \overset{\Delta}{=} Y_{ij}(Z^{\text{obs}}_{i1} \wedge Z^{\text{obs}}_{i2}),
\end{split}
\end{equation*}
where we abuse the notation and use subscripts $T$ and $C$ to denote the potential outcomes under the maximum and minimum of two observed doses within each matched pair, respectively.
\end{definition}

Write $\mathcal{F} = \{\mathbf{X}_{ij}, Y_{Tij}, Y_{Cij},~i = 1, \cdots, I,~j = 1, 2\}$, where $Y_{Tij}$ and $Y_{Cij}$ are defined in Definition \ref{def: define potential outcomes}, $\mathbf{Z}^{\text{obs}}_{\vee} = (Z^{\text{obs}}_{11} \vee Z^{\text{obs}}_{12}, \cdots, Z^{\text{obs}}_{I1} \vee Z^{\text{obs}}_{I2})$, and $\mathbf{Z}^{\text{obs}}_{\wedge} = (Z^{\text{obs}}_{11} \wedge Z^{\text{obs}}_{12}, \cdots, Z^{\text{obs}}_{I1} \wedge Z^{\text{obs}}_{I2})$. As always in randomization inference (\citealp{rosenbaum2002observational, rosenbaum2010design, Ding2016variation}), we condition on observed covariates, potential outcomes, and observed dose assignments, i.e., we do not model $\mathbf{X}$ or the potential outcomes, and rely on the treatment assignment mechanism to draw causal conclusions. The law that describes the treatment dose assignment in each matched pair $i$ is 
\begin{equation*}
    \pi_{i1} = P(Z_{i1} = Z^{\text{obs}}_{i1} \vee Z^{\text{obs}}_{i2}, Z_{i2} = Z^{\text{obs}}_{i1} \wedge Z^{\text{obs}}_{i2} \mid \mathcal{F}, \mathbf{Z}^{\text{obs}}_{\vee}, \mathbf{Z}^{\text{obs}}_{\wedge}),
\end{equation*}
and $\pi_{i2} = 1 - \pi_{i1}$. In an ideal randomized experiment, experimenters use physical randomization (e.g., coin flips) to ensure $\pi_{i1} = \pi_{i2} = 1/2$: for matched pair $i$ with two treatment doses $Z^{\text{obs}}_{i1}$ and $Z^{\text{obs}}_{i2}$, a fair coin is flipped; if the coin lands heads, the first unit is assigned $Z^{\text{obs}}_{i1}$ and the second unit $Z^{\text{obs}}_{i2}$, and vice versa if the coin lands tails. The design stage of an observational study aims to approximate this ideal (yet unattainable) hypothetical experiment by matching units with similar covariates $\mathbf{X}$ so that $\pi_{i1} \approx \pi_{i2}$ after matching. In this way, nonbipartite matching embeds observational data with a continuous treatment dose into a randomized experiment; this induced randomization scheme will serve as our ``reasoned basis" for inferring any causal effect including a dose-response relationship. As is always true with retrospective observational studies, a careful design may alleviate, but most likely never eliminate bias due to the residual imbalance in $\mathbf{X}$ or unmeasured confounding variables. The departure from randomization, i.e., $\pi_{i1} \neq \pi_{i2}$, is investigated via a sensitivity analysis (\citealp{rosenbaum1989sensitivity, rosenbaum2002observational, rosenbaum2010design}). 

\section{Randomization-based inference for a dose-response relationship}
\label{sec: randomization inference time-indep}
\subsection{Randomization inference for $\tau = \tau_0$ and $\beta = \beta_0$ in the dose-response kink model}
\label{subsec: randomization for fixed params}
Endowed with the randomization scheme induced by nonbipartite matching, we now turn to statistical inference. We first consider testing the dose-response kink model for a fixed $\tau = \tau_0$ and $\beta = \beta_0$ for all units, i.e.,
\begin{equation*}
\begin{split}
H_{0, \text{kink}}^{\tau_0, \beta_0}:~
    &Y_{ij}(z) = Y_{ij}(z^\ast),~\forall z\leq \tau_0, ~\text{and}\\
    &Y_{ij}(z) - Y_{ij}(\tau_0) = \beta_0(z - \tau_0),~\forall z > \tau_0,~\forall i,j.
\end{split}
\end{equation*}
Under $H_{0, \text{kink}}^{\tau_0, \beta_0}$, the entire dose-response relationship for subject $ij$ is known up to $Y_{ij}(z^\ast)$. Fortunately, we do observe one point on the dose-response curve, namely $Y_{ij}(Z^{\text{obs}}_{ij})$; hence, the entire dose-response curve for the subject $ij$ is fixed, and both potential outcomes $Y_{ij}(Z^{\text{obs}}_{i1} \wedge Z^{\text{obs}}_{i2})$ and $Y_{ij}(Z^{\text{obs}}_{i1} \vee Z^{\text{obs}}_{i2})$ can then be imputed for each unit $ij$. In matched pair $i$, for the unit with $Z_{ij} = Z^{\text{obs}}_{i1} \wedge Z^{\text{obs}}_{i2}$, the potential outcome under $Z^{\text{obs}}_{i1} \wedge Z^{\text{obs}}_{i2}$ is the observed outcome $Y^{\text{obs}}_{ij}$ and under $Z^{\text{obs}}_{i1} \vee Z^{\text{obs}}_{i2}$ is
\begin{equation}
    \label{eqn: impute kink 1}
    \begin{cases}
Y_{ij}^{\text{obs}},\quad & Z^{\text{obs}}_{i1} \vee Z^{\text{obs}}_{i2} \leq \tau_0; \\
Y_{ij}^{\text{obs}} + \beta_0 \times (Z^{\text{obs}}_{i1} \vee Z^{\text{obs}}_{i2} - \tau_0), & Z^{\text{obs}}_{i1} \wedge Z^{\text{obs}}_{i2} \leq \tau_0~\text{and}~Z^{\text{obs}}_{i1} \vee Z^{\text{obs}}_{i2} > \tau_0; \\
Y_{ij}^{\text{obs}} + \beta_0 \times (Z^{\text{obs}}_{i1} \vee Z^{\text{obs}}_{i2} - Z^{\text{obs}}_{i1} \wedge Z^{\text{obs}}_{i2}), & Z^{\text{obs}}_{i1} \wedge Z^{\text{obs}}_{i2} > \tau_0. \\
\end{cases}
\end{equation}
Analogously, for the unit with $Z_{ij} = Z^{\text{obs}}_{i1} \vee Z^{\text{obs}}_{i2}$, the potential outcome under $Z^{\text{obs}}_{i1} \vee Z^{\text{obs}}_{i2}$ is the observed outcome $Y^{\text{obs}}_{ij}$ and under $Z^{\text{obs}}_{i1} \wedge Z^{\text{obs}}_{i2}$ is
\begin{equation}
    \label{eqn: impute kink 2}
    \begin{cases}
Y_{ij}^{\text{obs}},\quad & Z^{\text{obs}}_{i1} \vee Z^{\text{obs}}_{i2} \leq \tau_0; \\
Y_{ij}^{\text{obs}} - \beta_0 \times (Z^{\text{obs}}_{i1} \vee Z^{\text{obs}}_{i2} - \tau_0), & Z^{\text{obs}}_{i1} \vee Z^{\text{obs}}_{i2} > \tau_0~\text{and}~Z^{\text{obs}}_{i1} \wedge Z^{\text{obs}}_{i2} \leq \tau_0; \\
Y_{ij}^{\text{obs}} - \beta_0 \times (Z^{\text{obs}}_{i1} \vee Z^{\text{obs}}_{i2} - Z^{\text{obs}}_{i1} \wedge Z^{\text{obs}}_{i2}), & Z^{\text{obs}}_{i1} \wedge Z^{\text{obs}}_{i2} > \tau_0. \\
\end{cases}
\end{equation}
Table \ref{tbl: science table impute} illustrates the imputation scheme by imputing the missing potential outcome for each subject under the null hypothesis $H_{0, \text{kink}}^{\tau_0, \beta_0}$ with $\tau_0 = 0.3$ and $\beta_0 = 1$.

\begin{table}[h]
\centering
\caption{\small Imputed science table when testing the dose-response kink model with $\tau_0 = 0.3$ and $\beta_0 = 1$. Two units in each pair $i$ are arranged so that $i1$ has a smaller dose and $i2$ a larger dose. For each unit, one and only one potential outcome is observed and the other one imputed under $H_{0,\text{kink}}^{\tau_0, \beta_0}$.}
\label{tbl: science table impute}
\resizebox{\textwidth}{!}{
\fbox{%
\begin{tabular}{cccccccc} \\[-0.8em]
 & & \multicolumn{2}{c}{Observe One Potential Outcome}
&\multicolumn{2}{c}{Imputed Potential Outcomes} \\ \\[-0.9em]  \cline{3-6} \cline{3-6}
\multirow{3}{*}{\begin{tabular}{c}Units\end{tabular}}  
&\multirow{3}{*}{\begin{tabular}{c}Observed\\Dose $Z^{\text{obs}}_{ij}$\end{tabular}}
&\multirow{3}{*}{\begin{tabular}{c}$Y_{ij}(Z^{\text{obs}}_{i1} \wedge Z^{\text{obs}}_{i2})$\end{tabular}} 
&\multirow{3}{*}{\begin{tabular}{c}$Y_{ij}(Z^{\text{obs}}_{i1} \vee Z^{\text{obs}}_{i2})$\end{tabular}} 
&\multirow{3}{*}{\begin{tabular}{c}$Y_{ij}(Z^{\text{obs}}_{i1} \wedge Z^{\text{obs}}_{i2})$\end{tabular}} 
&\multirow{3}{*}{\begin{tabular}{c}$Y_{ij}(Z^{\text{obs}}_{i1} \vee Z^{\text{obs}}_{i2})$\end{tabular}} 
\\ \\ \\[-0.9em]  \\ \hline \\[-0.9em]
$11$ & $0.2$ & $Y^{\text{obs}}_{11}$ & $\boldsymbol{?}$ & $Y^{\text{obs}}_{11}$ & $Y^{\text{obs}}_{11}$
\\
$12$ & $0.4$ & $\boldsymbol{?}$ &$Y^{\text{obs}}_{12}$ & $Y^{\text{obs}}_{12}$ & $Y^{\text{obs}}_{12}$
\\

$21$ & $0.9$  & $Y^{\text{obs}}_{21}$ & $\boldsymbol{?}$ & $Y^{\text{obs}}_{21}$ & $Y^{\text{obs}}_{21} + 0.3 \times (2.2 - 1)$
\\
$22$ & $2.2$  & $\boldsymbol{?}$ &$Y^{\text{obs}}_{22}$ & $Y^{\text{obs}}_{22} - 0.3\times (2.2 - 1)$ & $Y^{\text{obs}}_{22}$
\\

$31$ & $1.4$  & $Y^{\text{obs}}_{31}$ & $\boldsymbol{?}$ & $Y^{\text{obs}}_{31}$ & $Y^{\text{obs}}_{31} + 0.3 \times (1.9 - 1.4)$
\\
$32$ & $1.9$  & $\boldsymbol{?}$ &$Y^{\text{obs}}_{32}$ & $Y^{\text{obs}}_{32} - 0.3\times (1.9 - 1.4)$ & $Y^{\text{obs}}_{32}$
\\

$\vdots$ & $\vdots$ & $\vdots$ & $\vdots$ &$\vdots$ &$\vdots$\\

\multirow{2}{*}{\begin{tabular}{c}$I1$ \end{tabular}} 
&\multirow{2}{*}{\begin{tabular}{c}$Z_{I1}$ \end{tabular}}
& \multirow{2}{*}{\begin{tabular}{c}$Y^{\text{obs}}_{I1}$ \end{tabular}} 
& \multirow{2}{*}{\begin{tabular}{c}$\boldsymbol{?}$ \end{tabular}} 
&\multirow{2}{*}{\begin{tabular}{c}$Y^{\text{obs}}_{I1}$ \end{tabular}} 
& \multirow{2}{*}{\begin{tabular}{c}
Impute according \\ to scheme \eqref{eqn: impute kink 1}
\end{tabular}} 
\\ \\

\multirow{2}{*}{\begin{tabular}{c}$I2$ \end{tabular}} 
&\multirow{2}{*}{\begin{tabular}{c}$Z_{I2}$ \end{tabular}}
& \multirow{2}{*}{\begin{tabular}{c}$\boldsymbol{?}$ \end{tabular}} 
& \multirow{2}{*}{\begin{tabular}{c}$Y^{\text{obs}}_{I2}$ \end{tabular}} 
& \multirow{2}{*}{\begin{tabular}{c}Impute according \\ to scheme \eqref{eqn: impute kink 2}\end{tabular}} 
&\multirow{2}{*}{\begin{tabular}{c}$Y^{\text{obs}}_{I2}$ \end{tabular}} 
\\ \\
\end{tabular}}}
\end{table}

Let $ij'$ denote the unit with dose $Z^{\text{obs}}_{i1} \wedge Z^{\text{obs}}_{i2}$ in matched pair $i$, $\mathcal{Y}^\text{obs}_{\min}$ = $\{Y^{\text{obs}}_{ij'},~i = 1,\cdots,I\}$, and $\widehat{F}_{\min}(\cdot)$ the CDF of $\mathcal{Y}^\text{obs}_{\min}$. Analogously, let $ij''$ denote the unit with dose $Z^{\text{obs}}_{i1} \vee Z^{\text{obs}}_{i2}$ and $\mathcal{Y}^\text{obs}_{\max}$ = $\{Y^{\text{obs}}_{ij''},~i = 1,\cdots,I\}$. For each $Y^{\text{obs}}_{ij''} \in \mathcal{Y}^\text{obs}_{\max}$, define the transformed outcome $\widetilde{Y}^{\text{obs}}_{ij''}$ to be unit $ij'$'s potential outcome under the dose $Z^{\text{obs}}_{i1} \wedge Z^{\text{obs}}_{i2}$ according to \eqref{eqn: impute kink 2}. Let $\widetilde{\mathcal{Y}}^\text{obs}_{\max}$ = $\{\widetilde{Y}^{\text{obs}}_{ij''},~i = 1,\cdots,I\}$ denote the collection of transformed outcomes, and $\widehat{F}^{\text{tr}}_{\max}(\cdot)$ its CDF. The null hypothesis $H_{0, \text{kink}}^{\tau_0, \beta_0}$ can then be tested by comparing the following Kolmogorov-Smirnov-type (KS) test statistic 
\begin{equation}
    \label{eqn: KS test statistic}
    t_{\text{KS}}(\tau_0, \beta_0) = \sup_{y}\left|\widehat{F}_{\min}(y) - \widehat{F}^{\text{tr}}_{\max}(y)\right|
\end{equation}
evaluated at the observed data to a reference distribution generated based on the imputed science table (e.g., Table \ref{tbl: science table impute}) and enumerating all $2^I$ possible randomizations: within each matched pair $i$, unit $i1$ receives $Z^{\text{obs}}_{i1} \vee Z^{\text{obs}}_{i2}$ and exhibits $Y_{i1}^{\text{obs}} = Y_{i1}(Z^{\text{obs}}_{i1} \vee Z^{\text{obs}}_{i2})$ and $i2$ receives $Z^{\text{obs}}_{i1} \wedge Z^{\text{obs}}_{i2}$ and exhibits $Y^{\text{obs}}_{i2} = Y_{i2}(Z^{\text{obs}}_{i1} \wedge Z^{\text{obs}}_{i2})$, or unit $i1$ receives $Z^{\text{obs}}_{i1} \wedge Z^{\text{obs}}_{i2}$ and exhibits $Y_{i1}^{\text{obs}} = Y_{i1}(Z^{\text{obs}}_{i1} \wedge Z^{\text{obs}}_{i2})$ and $i2$ receives $Z^{\text{obs}}_{i1} \vee Z^{\text{obs}}_{i2}$ and exhibits $Y^{\text{obs}}_{i2} = Y_{i2}(Z^{\text{obs}}_{i1} \vee Z^{\text{obs}}_{i2})$. In principle, any test statistic can be combined with this randomization scheme to yield a valid test. We motivate the test statistic \eqref{eqn: KS test statistic} in Supplementary Material B. Note that when $\tau_0 = \infty$ or $\beta_0 = 0$, $H_{0, \text{kink}}^{\tau_0, \beta_0}$ reduces to the following causal null hypothesis:
\begin{equation*}
    H_{0, \text{null}}: Y_{ij}(z) = Y_{ij}(z^\ast),~\forall z \in \mathcal{Z}, \forall i = 1, \cdots, I, j = 1, 2,
\end{equation*}
and the developed procedure can be used to test $H_{0, \text{null}}$.

\begin{figure}[ht]
   \centering
     \subfloat{\includegraphics[width = 0.49\columnwidth]{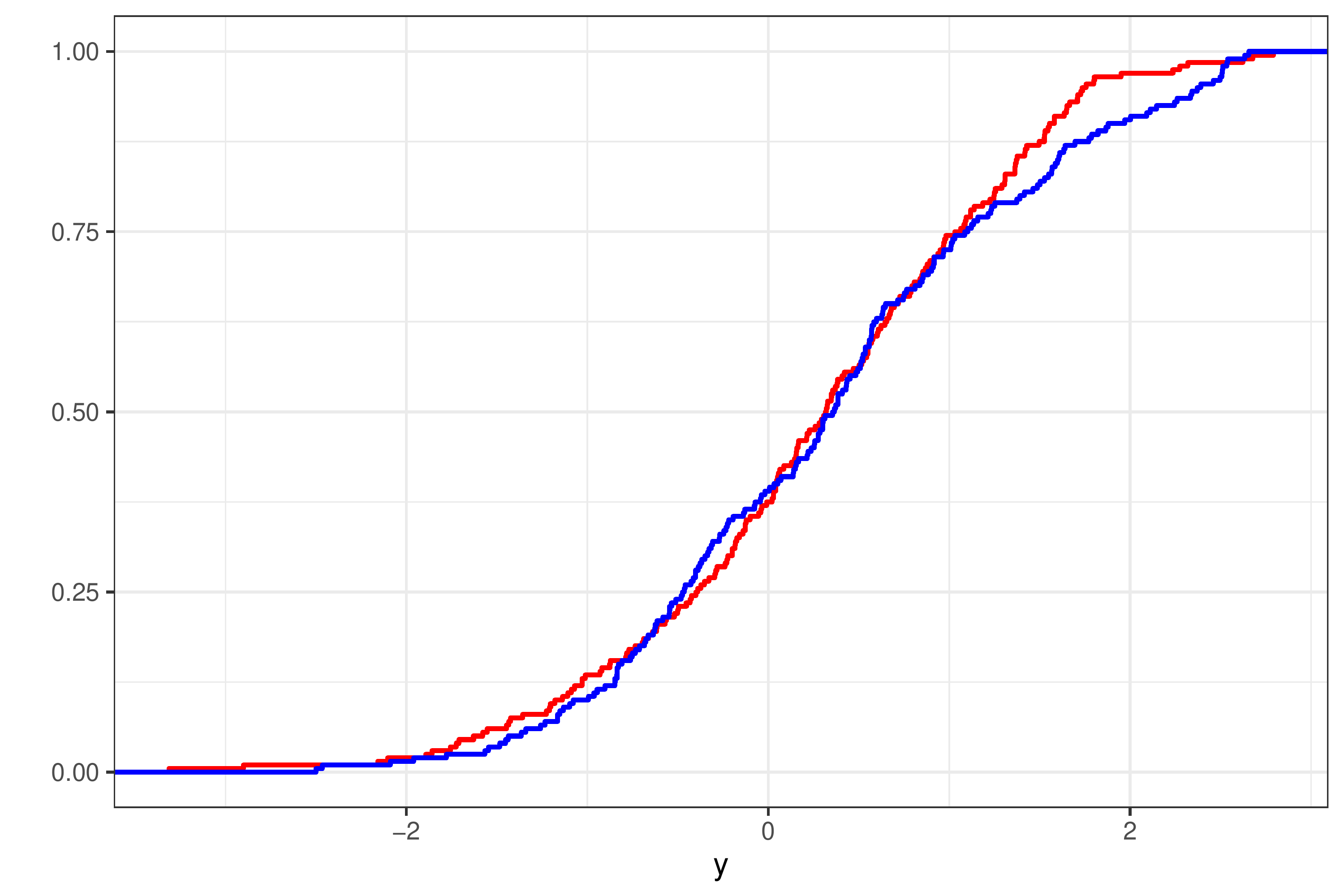}}
     \subfloat{\includegraphics[width = 0.49\columnwidth]{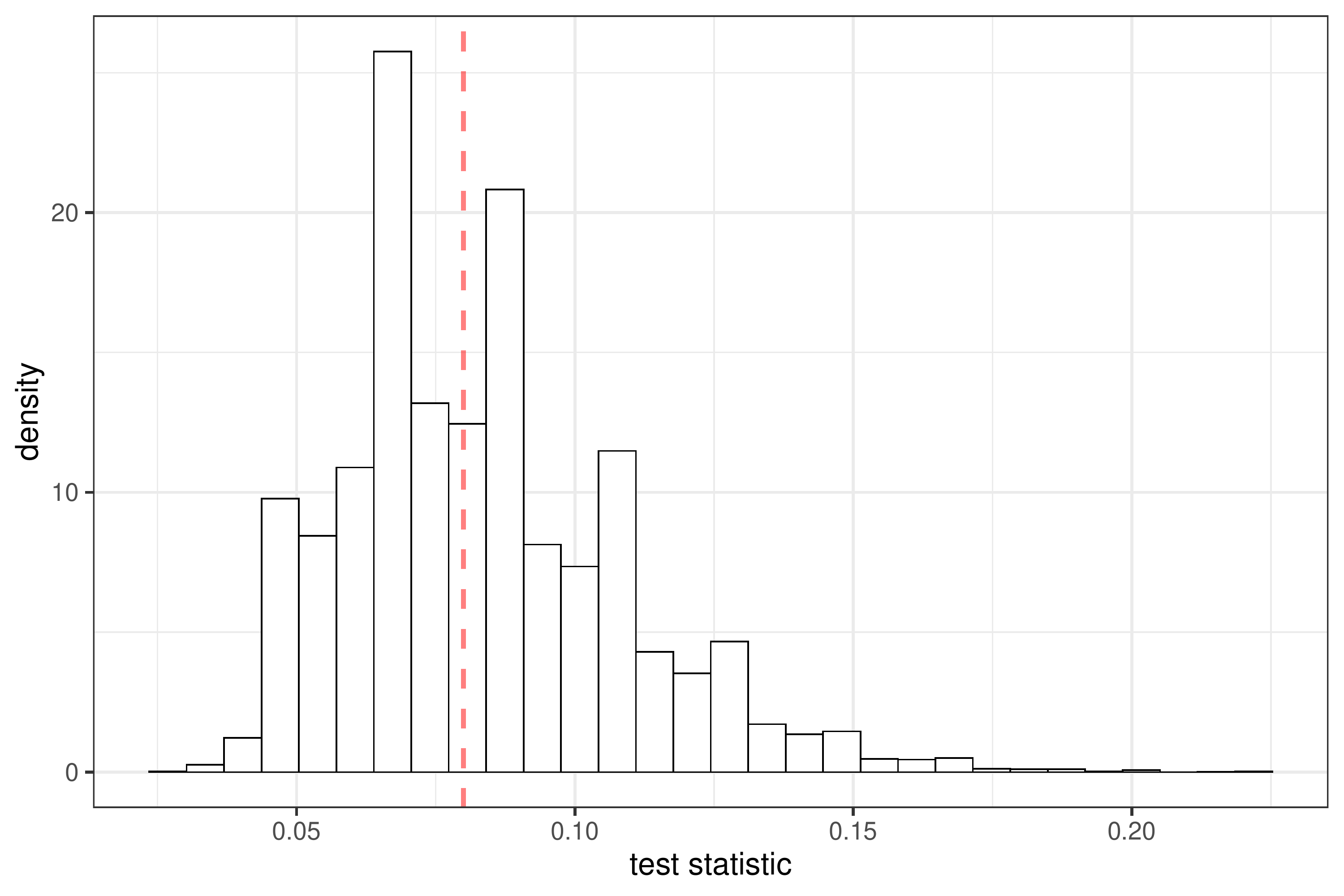}}
     \caption{\small An illustrative example. $I = 200$, $\tau = 1$, and $\beta = 0.5$. We test the null hypothesis $H^{\tau_0,\beta_0}_{0,\text{kink}}$ with $\tau_0 = 1$ and $\beta_0 = 0.5$. The left panel plots $\widehat{F}_{\min}(y)$, the empirical CDF of $\mathcal{Y}^\text{obs}_{\min}$ (blue) and $\widehat{F}^{\text{tr}}_{\max}(y)$, the empirical CDF of the transformed outcomes $\widetilde{\mathcal{Y}}^\text{obs}_{\max}$ (red). The test statistic $t_{\text{KS}}(1, 0.5)$ evaluated at the observed data is $0.08$. The right panel plots the exact reference distribution of the test statistic given the sample and under the null hypothesis. The reference distribution is generated using $100,000$ Monte Carlo draws from the $2^{200}$ randomization configurations. The red dashed line plots the position of the observed test statistic. The exact p-value in this case is $0.445$.}
\label{fig: illustrate KS test}
\end{figure}

We illustrate the procedure using the following example. We generate $I = 200$ matched pairs of $2$ units, each with $Z^{\text{obs}}_{ij} \sim \text{Unif}[0, 4]$, $Y_{ij}(0) \sim \text{Normal}(0, 1)$, and $Y^{\text{obs}}_{ij} = Y_{ij}(Z^{\text{obs}}_{ij})$ follows Model \eqref{eqn: dose-response kink model} with $\tau = 1$ and $\beta = 0.5$. We test the null hypothesis $H^{\tau_0,\beta_0}_{0,\text{kink}}$ with $\tau_0 = 1$ and $\beta_0 = 0.5$ using the test statistic \eqref{eqn: KS test statistic}. The left panel of Figure \ref{fig: illustrate KS test} plots the empirical distribution $\widehat{F}_{\min}(y)$ (blue) and $\widehat{F}^{\text{tr}}_{\max}(y)$ (red), and $t_{\text{KS}}(1, 0.5) = 0.08$ for the observed data. Instead of enumerating all $2^I = 2^{200}$ possible treatment dose assignments, we draw with replacement $100,000$ samples from all $2^{200}$ possible configurations. The right panel of Figure \ref{fig: illustrate KS test} plots the reference distribution based on these $100,000$ samples. Such a ``sampling with replacement" strategy is referred to as a ``modified randomization test" in the literature (\citealp{dwass1957modified, pagano1983obtaining}) and known to still preserve the level of the test. In this way, a p-value equal to $0.445$ is obtained in this simulated dataset and the null hypothesis $H_{0, \text{kink}}^{\tau_0, \beta_0}$ with $\tau_0 = 1$ and $\beta_0 = 0.5$ is not rejected. The p-value is exact as the procedure does not resort to any asymptotic theory and works in small samples.

\subsection{Testing the dose-response kink model}
Let $H^K_0$ denote a composite hypothesis that is equal to the union of $H_{0, \text{kink}}^{\tau_0, \beta_0}$ over all $\tau = \tau_0$ and $\beta = \beta_0$, i.e.,
\begin{equation*}
    H^K_0 = \bigcup_{\tau_0, \beta_0} H_{0, \text{kink}}^{\tau_0, \beta_0}.
\end{equation*}
In other words, the activation dose $\tau$ and the slope $\beta$ are nuisance parameters to be taken into account. One strategy testing $H_0^K$ is to take the supremum p-value over the entire range of $(\tau, \beta)$; another commonly used strategy (\citealp{berger1994p}) is to first construct a confidence set around $(\tau, \beta)$ and then take the supremum p-values over the $(\tau, \beta)$ values in this confidence set. This latter strategy is particularly useful when the treatment dose and/or the outcome of interest are not bounded so that $\tau$ and $\beta$ are not bounded; for some applications in the causal inference literature, see \citet{nolen2011randomization}, \citet{Ding2016variation}, and \citet{zhang2021bridging}. In Supplementary Material C, we discuss how to construct a bounded level-$\gamma$ confidence set for $(\tau, \beta)$ based on inverting a variant of the Wilcoxon rank sum test statistic and its properties.

\begin{figure}[h]
    \centering
    \includegraphics[width=0.7\columnwidth]{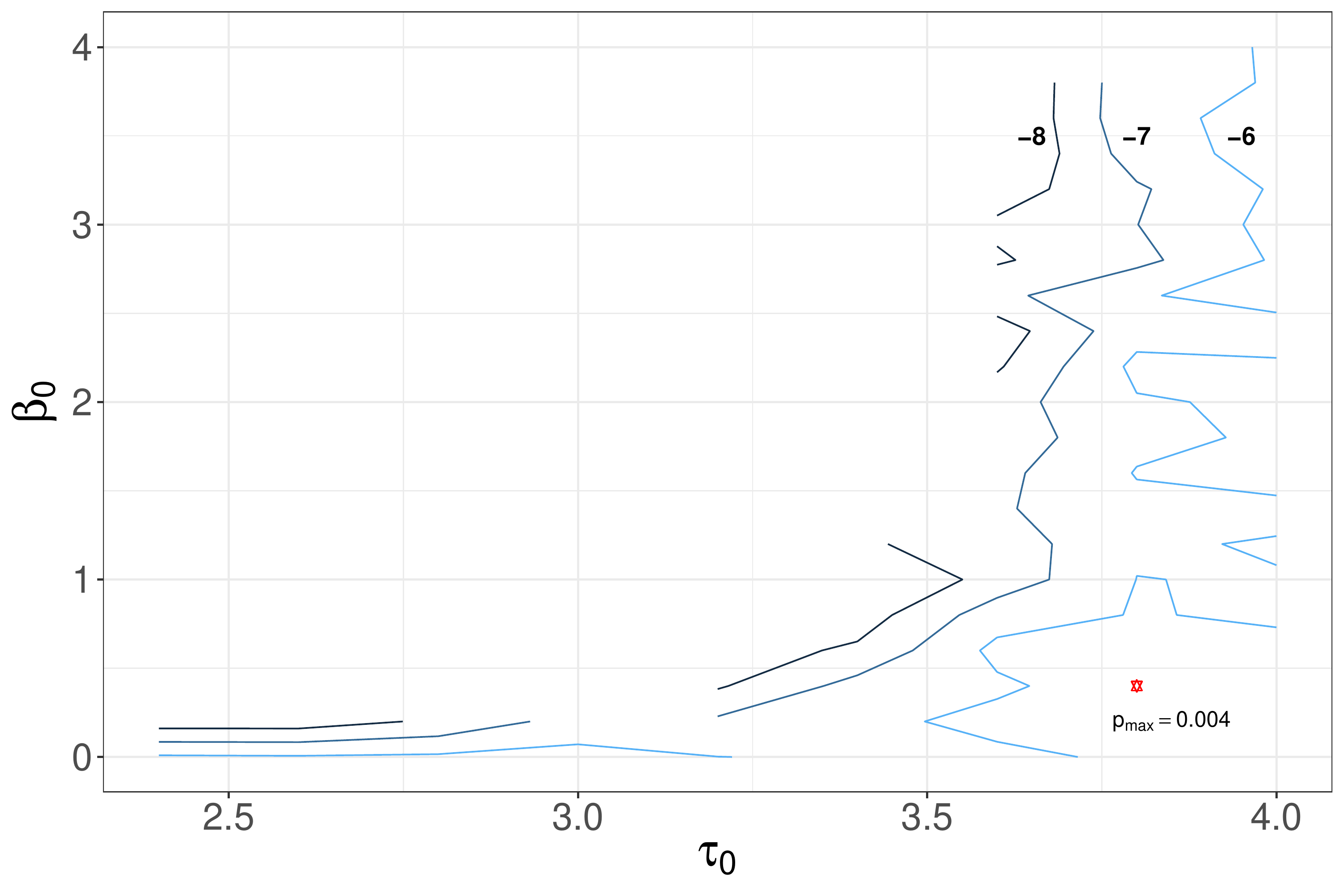}
    \caption{\small The probability contour plot (in log scale) against values of $\tau_0$ and $\beta_0$. The true dose-response model is $Y_{ij}(z) = Y_{ij}(0) + 2\cdot\mathbbm{1}\{0 \leq z \leq 1\} + 1\cdot \mathbbm{1}\{1 < z \leq 4\}$. We let $Y_{ij}(0) \sim N(0, 1)$ and $I = 200$. We test $ H_{0, \text{kink}}^{\tau_0, \beta_0}$ and plot the p-value in log scale against $\tau_0$ and $\beta_0$ values. The maximum p-value is obtained at $\tau_0 = 3.8$ and $\beta_0 = 0.4$ and equal to $0.004$. The null hypothesis $H_0^K$ is hence rejected at level $0.05$ for this simulated dataset.}
    \label{fig: illustrate testing kink model}
\end{figure}

Being able to reject $H_0^K$ suggests evidence against the postulated dose-response relationship; otherwise, the model is deemed sufficient to characterize the dose-response relationship for the data at hand. We illustrate the procedure using the following example. We generate $I = 200$ matched pairs of $2$ units with $Z^{\text{obs}}_{ij} \sim \text{Unif}[0, 4]$, $Y_{ij}(0) \sim \text{Normal}(0, 1)$, and $Y^{\text{obs}}_{ij} = Y_{ij}(Z^{\text{obs}}_{ij}) = Y_{ij}(0) + 2\cdot\mathbbm{1}\{0 \leq Z^{\text{obs}}_{ij} \leq 1\} + 1\cdot \mathbbm{1}\{1 < Z^{\text{obs}}_{ij} \leq 4\}$. Figure \ref{fig: illustrate testing kink model} plots the p-values in log scale against $\tau_0$ and $\beta_0$. The maximum p-value is obtained at $\tau_0 = 3.8$ and $\beta_0 = 0.4$ and equal to $0.004$. The null hypothesis $H_0^K$, i.e., the dose-response relationship follows a kink model, can be rejected at level $0.05$ for this simulated dataset.

\subsection{Testing any structured dose-response model}
\label{subsec: test structured dose-response model}
Our discussion above suggests a general model-free, randomization-based framework to test any structured dose-response relationship. Here, we say a dose-response relationship is ``structured" if it is characterized by a few structural parameters. Consider the following structured dose-response relationship model:
\begin{equation*}
   H_0^{\text{dose-response}}: Y_{ij}(z) - Y_{ij}(z^\ast) \triangleq f(z; z^\ast,\boldsymbol{\theta}) = 0,~\forall i = 1,\cdots,I,~j = 1,2,~\text{for some}~\boldsymbol{\theta},
\end{equation*}
where $z^\ast \in \mathcal{Z}$ is a reference dose, and $f(\cdot~; z^\ast, \boldsymbol{\theta})$ is a univariate function that satisfies $f(z^\ast; z^\ast, \boldsymbol\theta) = 0$ and is parametrized by a $p$-dimensional vector of structural parameters $\boldsymbol{\theta} \in \mathbb{R}^p$. Algorithm \ref{alg} summarizes a general procedure testing $H_0^{\text{dose-response}}$ at level $\alpha$. In Supplementary Material D, we briefly discuss and illustrate how to sequentially test a few dose-response relationships ordered in their model complexity.

\begin{algorithm} 
\SetAlgoLined
\caption{\small Randomization Inference for a Dose-Response Relationship: Pseudo Algorithm} \label{alg}
\vspace*{0.12 cm}
\KwIn{$I$ matched pairs after nonbipartite matching and a dose-response relationship model $H_0^{\text{dose-response}}: Y_{ij}(z) - Y_{ij}(z^\ast) \triangleq f(z; z^\ast,\boldsymbol{\theta}) = 0,~\forall i,j,~\text{for some}~\boldsymbol\theta$}
\vspace*{0.12 cm}
\begin{enumerate}
    \item Construct $\text{CI}_{\boldsymbol{\theta}}$, a level-$\gamma$ confidence set for the structural parameter $\boldsymbol{\theta}$;
    \item For each $\boldsymbol{\theta}_0 \in \text{CI}_{\boldsymbol{\theta}}$, do the following steps:

\begin{enumerate}
        \item Compute the test statistic $t^{\text{obs}}$. For each unit $ij$ with $Z_{ij} = Z^{\text{obs}}_{i1} \vee Z^{\text{obs}}_{i2}$, i.e., the unit with maximum dose in each matched pair $i$, define the following transformed outcome
    \begin{equation}
    \widetilde{Y}^{\text{obs}}_{ij} = Y^{\text{obs}}_{ij} - f(Z^{\text{obs}}_{i1} \vee Z^{\text{obs}}_{i2}; z^\ast, \boldsymbol{\theta}_0) + f(Z^{\text{obs}}_{i1} \wedge Z^{\text{obs}}_{i2}; z^\ast, \boldsymbol{\theta}_0).
    \end{equation}
    Let $\widehat{F}^{\text{tr}}_{\max}(\cdot)$ denote the empirical CDF of $\{\widetilde{Y}^{\text{obs}}_{ij},~i = 1,\cdots,I\}$ and $\widehat{F}_{\min}(\cdot)$ the empirical CDF of the collection of units $ij$ with $Z_{ij} = Z^{\text{obs}}_{i1} \wedge Z^{\text{obs}}_{i2}$. Calculate \[
    t^{\text{obs}} = \sup_{y}\left|\widehat{F}_{\min}(y) - \widehat{F}^{\text{tr}}_{\max}(y)\right|;    
    \]
    
    \item Impute the science table. For each unit $ij$ with $Z_{ij} = Z^{\text{obs}}_{i1} \wedge Z^{\text{obs}}_{i2}$, impute $Y_{ij}(Z^{\text{obs}}_{i1} \wedge Z^{\text{obs}}_{i2}) = Y_{ij}^{\text{obs}}$ and \[Y_{ij}(Z^{\text{obs}}_{i1} \vee Z^{\text{obs}}_{i2}) = Y_{ij}^{\text{obs}} + f(Z^{\text{obs}}_{i1} \vee Z^{\text{obs}}_{i2}; z^\ast, \boldsymbol{\theta}_0) - f(Z^{\text{obs}}_{i1} \wedge Z^{\text{obs}}_{i2}; z^\ast, \boldsymbol{\theta}_0);
        \]for each unit $ij$ with $Z_{ij} = Z^{\text{obs}}_{i1} \vee Z^{\text{obs}}_{i2}$, impute $Y_{ij}(Z^{\text{obs}}_{i1} \vee Z^{\text{obs}}_{i2}) = Y_{ij}^{\text{obs}}$ and \[
        Y_{ij}(Z^{\text{obs}}_{i1} \wedge Z^{\text{obs}}_{i2}) = Y_{ij}^{\text{obs}} - f(Z^{\text{obs}}_{i1} \vee Z^{\text{obs}}_{i2}; z^\ast, \boldsymbol{\theta}_0) + f(Z^{\text{obs}}_{i1} \wedge Z^{\text{obs}}_{i2}; z^\ast, \boldsymbol{\theta}_0);
    \]
    \item Generate a reference distribution. Sample with replacement $\text{MC} = 100,000$ dose assignment configurations from the $2^I$ possible configurations. For each sampled dose assignment configuration $\widetilde{\boldsymbol{Z}}_k$, calculate $\widetilde{t}^{(k)}_{\text{KS}}(\boldsymbol\theta_0)$ according to Step (a). Let $\widetilde{F}_{\boldsymbol{\theta}_0}$ denote the distribution of $\{\widetilde{t}^{(k)}_{\text{KS}}(\boldsymbol{\theta}_0),~k = 1, 2, \cdots, \text{MC}\}$;
    
    \item Compute the p-value $p_{\boldsymbol\theta_0}$ by comparing $t^{\text{obs}}$ to the reference distribution $\widetilde{F}_{\boldsymbol{\theta_0}}$, i.e.,
    \[
    p_{\boldsymbol\theta_0} = \frac{1}{\text{MC}} \sum_{k = 1}^{\text{MC}} \mathbbm{1}\left\{\Tilde{t}^{(k)}_{\text{KS}}(\boldsymbol{\theta}_0) \geq t^{\text{obs}}\right\};
    \]
    \end{enumerate} 
    \vspace{-0.2 cm}
\item Let $p_{\max} = \sup_{\boldsymbol{\theta}\in\text{CI}_{\boldsymbol\theta}} p_{\boldsymbol\theta}$ and reject the null hypothesis $H_0^{\text{dose-response}}$ at level $\alpha$ if $p_{\max} + \gamma \leq \alpha$.
\end{enumerate}
\end{algorithm}

\section{Relaxing the SUTVA: dose-response relationship under interference}
\label{sec: interference}
\subsection{Potential outcomes under interference}
We relax the stable unit treatment value assumption in this section and consider inference for a structured dose-response relationship under interference. To this end, we collect the treatment doses of all study units in our matched-pair design and use $\Vec{\mathbf{Z}} = (Z_{11}, Z_{12}, \cdots, Z_{I1}, Z_{I2})$ to represent the treatment dose configuration with $\Vec{\mathbf{z}}$ being its realization. We further let $\Vec{\mathbf{Z}}^{\text{obs}}$ denote the observed treatment dose configuration of all $2I$ study units and
\begin{equation}
\label{eqn: potential outcome under interference}
 Y_{ij}(\Vec{\mathbf{Z}}) := Y_{ij}(Z_{11}, \cdots, Z_{I2})
\end{equation}
unit $ij$'s potential outcome that is random only through the randomness in the treatment dose configuration $\Vec{\mathbf{Z}}$. The SUTVA states that for all pairs of $\Vec{\mathbf{z}}$ and $\Vec{\mathbf{z}}'$, $z_{ij} = z'_{ij}$ implies $Y_{ij}(\Vec{\mathbf{z}}) = Y_{ij}(\Vec{\mathbf{z}}')$; in other words, $Y_{ij}(\Vec{\mathbf{Z}})$ depends on $\Vec{\mathbf{Z}}$ only through its dependence on $Z_{ij}$. 

Definition \eqref{eqn: potential outcome under interference} is in a most general form and useful when the scientific interest lies in testing the null hypothesis of no direct or spillover effect under \emph{arbitrary} interference pattern. To further explore the dose-response relationship in the presence of the spillover effect, researchers need to model the local interference structure possibly based on units' spatial relationship (e.g., closeness of counties in our case study). To this end, we assume study units are connected through an undirected network with a symmetric, $2I \times 2I$ adjacency matrix $\mathbf{G}$. Matrix $\mathbf{G}$ has its rows and columns arranged in the order corresponding to unit $11, 12, \cdots, I1, I2$ after nonbipartite matching. If unit $ij$ and $i'j'$ are connected, then the corresponding entry in $\mathbf{G}$ is equal to $1$ and otherwise $0$. The diagonal entries of $\mathbf{G}$ are defined to be $0$.

Our reasoned basis for testing any causal null hypothesis \emph{under interference} will still be the randomization scheme endowed by the nonbipartite matching. We have two goals. First, we show that the test developed for $H_{0, \text{null}}$ under the SUTVA remains a valid level-$\alpha$ test for a null hypothesis of no direct or spillover effect under arbitrary interference pattern. Second, we relax the dose-response relationship $H_{0, \text{kink}}$ by modeling various forms of local interference pattern using the adjacency matrix $\mathbf{G}$.

\subsection{No direct or spillover effect}
\label{subsec: no direct or spillover}
Following \citet{rosenbaum2007interference, bowers2013reasoning, athey2018exact}, a null hypothesis of no direct or spillover effect states that
\begin{equation*}
   H_{0, \text{direct or spillover}}: Y_{ij}(\Vec{\mathbf{z}}) = Y_{ij}(\Vec{\mathbf{z}}'),~\forall i = 1, \cdots, I, j = 1, 2,
\end{equation*}
and all pairs of treatment dose configurations of $2I$ study units $\Vec{\mathbf{z}}$ and $\Vec{\mathbf{z}}'$. Under $H_{0, \text{direct or spillover}}$, the unit-level potential outcome of each study unit under any treatment dose configuration $\Vec{\mathbf{z}}$ can still be imputed; in fact, $Y_{ij}(\Vec{\mathbf{z}}) = Y_{ij}(\Vec{\mathbf{Z}}^{\text{obs}})$ for any $\Vec{\mathbf{z}}$. Any test statistic (e.g., the Kolmogorov-Smirnov statistic used in Algorithm \ref{alg}) that depends on units' potential outcomes (possibly under interference) is random only through its dependence on the treatment dose configurations of all study units; therefore, the null distribution of the test statistic can again be inferred by enumerating $2^I$ different configurations of $\Vec{\mathbf{Z}}$ as discussed in Section \ref{sec: randomization inference time-indep}. In other words, the testing procedure for $H_{0, \text{null}}$ is still exact and has correct level for testing $H_{0, \text{direct or spillover}}$. Moreover, since $H_{0, \text{direct or spillover}}$ does \emph{not} impose any interference pattern, rejecting $H_{0, \text{null}}$ implies rejecting $H_{0, \text{direct or spillover}}$ under \emph{arbitrary} interference pattern.

\subsection{Dose-response relationship under local interference modeling}
\label{subsec: dose-response under interference}
Testing the null hypothesis is often regarded a starting point of causal analysis (\citealp{imbens2015causal}). Next, we build up a causal hypothesis regarding a dose-response relationship allowing for local interference. Our construction is guided by the following general principles adapted from the literature on interference (\citealp{hong2006evaluating, bowers2013reasoning, athey2018exact})
\begin{description}
\item[\textbf{Principle I:}] The total effect of treatment dose configuration $\Vec{\mathbf{z}}$ compared to a reference dose configuration $\Vec{\mathbf{z}}^\ast$ can be decomposed into a dose-response direct effect due to $ij$'s own treatment dose $z_{ij}$ and a spillover effect due to other study units' treatment doses so that $Y_{ij}(\vec{\mathbf{z}}) - Y_{ij}(\Vec{\mathbf{z}}^\ast) = f(z_{ij}; z^\ast_{ij}, \boldsymbol{\theta}) + g(\vec{\mathbf{z}}_{-ij}; \vec{\mathbf{z}}_{-ij}^\ast)$ where $f(z_{ij}; z^\ast_{ij}, \boldsymbol{\theta})$ is a dose-response direct effect described in Section \ref{sec: randomization inference time-indep}, $\vec{\mathbf{z}}_{-ij}$ (resp. $\vec{\mathbf{z}}_{-ij}^\ast$) treatment doses (resp. reference treatment doses) of all study units except $ij$, and $g(\cdot)$ a function modeling the spillover effect. For a binary treatment, $\vec{\mathbf{z}}^\ast = \vec{\mathbf{0}}$ is referred to as a uniformity trial (\citealp{rosenbaum2007interference}).
\item[\textbf{Principle II:}]  The spillover effect depends only on the aggregate, excess treatment doses of $ij$'s neighbors with respect to the reference dose configuration so that $Y_{ij}(\vec{\mathbf{z}}) - Y_{ij}(\Vec{\mathbf{z}}^\ast) = f(z_{ij}; z^\ast_{ij}, \boldsymbol{\theta}) + g(\langle\vec{\mathbf{z}} - \vec{\mathbf{z}}^\ast, \mathbf{G}_{ij,\bigcdot}\rangle)$ where $\mathbf{G}_{ij, \bigcdot}$ is the $ij$-th row of the adjacency matrix $\mathbf{G}$.
\item[\textbf{Principle III:}] The spillover effect is always dominated by the dose-response direct effect in the sense that 
  \begin{equation}
  \label{eqn: spillover constraint}
       \|\mathbf{G}_{ij, \bigcdot}\|^{-1}_0\cdot \langle\vec{\mathbf{z}} - \vec{\mathbf{z}}^\ast, \mathbf{G}_{ij,\bigcdot}\rangle \leq z_{ij} - z^\ast_{ij} \quad\text{implies}\quad g(\langle\vec{\mathbf{z}} - \vec{\mathbf{z}}^\ast, \mathbf{G}_{ij,\bigcdot}\rangle) \leq f(z_{ij}; z^\ast_{ij}, \boldsymbol{\theta}). 
  \end{equation}
  One simple modeling strategy of $g(\langle\vec{\mathbf{z}} - \vec{\mathbf{z}}^\ast, \mathbf{G}_{ij,\bigcdot}\rangle)$ that satisfies \eqref{eqn: spillover constraint} is to scale the magnitude of the dose-response direct effect towards zero.
\end{description}

To illustrate the three principles above, we consider a concrete example of causal hypothesis under local interference. We consider a causal null hypothesis that states that the direct effect is proportional to the dose difference, i.e., $f(z_{ij}; z^\ast_{ij}, \boldsymbol\theta) = \beta(z_{ij} - z^\ast_{ij})$. We then model the local interference pattern by scaling the direct effect using a logistic function so that $g(\langle\vec{\mathbf{z}}- \vec{\mathbf{z}}^\ast, \mathbf{G}_{ij,\bigcdot}\rangle) = C \times f(z_{ij}; z^\ast_{ij}, \mathbf{\theta})$ with $C = 1/(1 + \exp\{-k(\langle\vec{\mathbf{z}} - \vec{\mathbf{z}}^\ast, \mathbf{G}_{ij,\bigcdot}\rangle - s)\})$. According to this specification, the spillover effect modeled by $g(\langle\vec{\mathbf{z}}-\vec{\mathbf{z}}^\ast, \mathbf{G}_{ij,\bigcdot}\rangle)$ trivially satisfies the third principle above as the multiplication factor $C$ is always upper bounded by $1$. The causal null hypothesis then becomes
\begin{equation*}
    H_{0, \text{interference}}: Y_{ij}(\Vec{\mathbf{z}}) - Y_{ij}(\Vec{\mathbf{z}}^\ast) = \beta(z_{ij} - z^\ast_{ij}) \cdot\left\{ 1 + \frac{1}{1 + \exp\{-k(\langle\vec{\mathbf{z}}-\vec{\mathbf{z}}^\ast, \mathbf{G}_{ij,\bigcdot}\rangle - s)\}}\right\}.
\end{equation*}

Statistical inference in the presence of interference parameters $(k, s)$ depends on one's perspective on $(k, s)$ (\citealp{bowers2013reasoning}). Inference may proceed by regarding interference parameters as sensitivity parameters and researchers could report how confidence sets of the dose-response relationship parameters in the direct effect (e.g., $\beta$ in $H_{0, \text{interference}}$) change as interference parameters change. For fixed interference parameters $(k_0, s_0)$, we can test $\beta = \beta_0$ in $H_{0, \text{interference}}$ by imputing potential outcomes for each study unit and each of the $2^I$ treatment dose configurations $\vec{\mathbf{Z}}$ under $H_{0, \text{interference}}$, choosing a test statistic $t(\mathbf{Y}(\vec{\mathbf{Z}}), \vec{\mathbf{Z}})$ that is a function of potential outcomes of all study units $\mathbf{Y}(\vec{\mathbf{Z}})$ and random only via its dependence on $\vec{\mathbf{Z}}$, generating the randomization-based reference distribution of $t(\mathbf{Y}(\vec{\mathbf{Z}}), \vec{\mathbf{Z}})$, and comparing the observed test statistic $t(\mathbf{Y}(\vec{\mathbf{Z}}^{\text{obs}}), \vec{\mathbf{Z}}^{\text{obs}})$ to this reference distribution.

\section{Extension to longitudinal studies with a time-varying treatment}
\label{sec: extension to longitudinal studies}
\subsection{Treatment dose trajectory and potential outcome trajectory}
\label{subsec: extension to longitudinal setup}
In our application, the treatment dose evolves over time and the public-health-related outcomes, e.g., county-level COVID-19 related death toll, may depend on the treatment dose trajectory. We first consider the no-interference case. Let $t_0$ denote a baseline period and $t_1, t_2, \cdots, t_i, \cdots, T$ subsequent treatment periods. Fix $t_0 \leq t_i \leq t_j$ and let $\mathcal{Z}$ be the set of all possible treatment doses at each time point. Let 
\begin{equation*}
    \mathbf{Z}_{t_i: t_j} = (Z_{t_i}, Z_{t_i + 1}, \cdots, Z_{t_j}) \in \underbrace{\mathcal{Z} \times \cdots \times \mathcal{Z}}_{t_j - t_i + 1}
\end{equation*}
denote the random treatment dose trajectory of one study unit from $t_i$ to $t_j$ (\citealp{robins1986new, bojinov2019time}), $\mathbf{z}_{t_i: t_j}$ one realization of $\mathbf{Z}_{t_i: t_j}$, and $\mathbf{Z}^{\text{obs}}_{n, t_i: t_j} = (Z^{\text{obs}}_{n, t_i}, \cdots, Z^{\text{obs}}_{n, t_j})$
the observed treatment dose trajectory of unit $n$ from $t_i$ to $t_j$. In our application, $t_0$ denotes the start of the phased reopening and $\mathbf{Z}_{t_i: t_j}$ the trajectory of daily percentage change in total distance traveled from $t_i$ to $t_j$. We are interested in the effect of a sustained period of treatment on some future outcome. We assume that the treatment dose at time $t$ temporally precedes the outcome at time $t$. Fix a time $t$ and let $Y_{n, t}(\mathbf{z}_{t_0:t}) = Y_{n, t}(z_{t_0}, z_{t_1}, \cdots, z_{t})$
denote the potential outcome of unit $n$ at time $t$ under the treatment dose trajectory $\mathbf{Z}_{n; t_0:t} = \mathbf{z}_{t_0:t}$. We assume consistency so that $Y^{\text{obs}}_{n, t} = Y_{n, t}(\mathbf{Z}^{\text{obs}}_{n; t_0:t})$. Finally, we let $\mathbf{Y}_{n; t_i: t_j}(\mathbf{z}_{t_0:t_j})$ denote unit $n$'s potential outcome trajectory from time $t_i$ to $t_j$ under the treatment dose trajectory $\mathbf{z}_{t_0: t_j}$.




\subsection{Covariate history and sequential randomization assumption}


One unique feature of longitudinal data is that the observed outcome trajectory up to time $t - 1$ may confound the treatment dose at time $t$; this is particularly true in our application: if the COVID-19 related case and death numbers were high during the last week in a county, then residents may be more wary of the disease and reduce social mobility this week. Following the literature on longitudinal studies, we let $\overline{L}_{n, t}$ denote the time-dependent covariate process of unit $n$ up to but not including time $t$; $\overline{L}_{n, t}$ contains both time-independent covariates $\bf X_n$ and time-dependent covariates like the observed outcomes $\{Y^{\text{obs}}_{n, t_0}, Y^{\text{obs}}_{n, t_1}, \cdots, Y^{\text{obs}}_{n, t - 1}\}$. We further assume the sequential randomization assumption (SRA) (\citealp{robins1997marginal}), which states that conditional on the treatment history up to time $t - 1$ and covariate process up to time $t$, the treatment dose assignment at time $t$ is independent of the potential outcome trajectories, i.e.,
\begin{equation*}
\label{eqn: SRA}
    \mathbf{Y}_{n; t_0: T}(\mathbf{z}_{t_0: T}) \indep Z_{n, t} \mid \mathbf{Z}_{n; t_0: t - 1} = \mathbf{z}_{n; t_0: t - 1}, \overline{L}_{n, t}, ~\forall \mathbf{z}_{t_0: T}.
\end{equation*}
This assumption holds if residents' adopting the social distancing measures at time $t$ depends on (1) their history of adopting social distancing measures, (2) time-independent covariates, and (3) observed daily COVID-19 related case numbers and death toll up to time $t - 1$. See also \citet{mattei2019bayesian} for a relaxed version of this assumption.

\subsection{Cumulative treatment dose, $\mathcal{W}$-equivalence, and dose-response relationship in a longitudinal setting} 
\label{subsec: CD, W-equivalence, dose-response in longitudinal setting}
One general recipe for drawing causal inference from longitudinal data is to model the marginal distribution of the counterfactual outcomes $Y_{n, t}(\mathbf{z}_{t_0: t})$, or the marginal joint distribution of $\mathbf{Y}_{n; t_i: t_j}(\mathbf{z}_{t_0: t})$, as a function of the treatment trajectory and baseline covariates; see \citet{robins1986new, robins1994correcting,robins1999estimation,robins2000} for seminal works. For example, one simplest model may state that $N$ units are i.i.d. samples from a superpopulation such that the counterfactual mean of the outcome at time $t$ depends on the treatment dose trajectory and the time-independent covariates $\boldsymbol X$ through a known functional form $g(\cdot)$, i.e., $\mathbb{E}[Y_{t}(\mathbf{Z}_{t_0: t}) \mid \boldsymbol X] = g(\mathbf{Z}_{t_0: t}, \boldsymbol X; \boldsymbol{\beta})$, and the interest lies in efficient estimation of the structural parameters $\boldsymbol{\beta}$.

In the infectious disease context, modeling the potential outcomes is a daunting task and our interest here lies in testing a structural dose response relationship in a less model-dependent way. To proceed, we generalize the notion of ``dose" from the static to longitudinal setting. Consider the following weighted difference between two treatment dose trajectories $\mathbf{z}_{t_i: t_j}$ and $\mathbf{z}'_{t_i: t_j}$:
\begin{equation}
\label{eqn: weighted l1 norm}
    \left\|\mathbf{z}_{t_i: t_j} - \mathbf{z}'_{t_i: t_j}\right\|_{\mathcal{W}} = \sum_{t_i \leq t' \leq t_j} w(t')\cdot (z_{t'} - z'_{t'}),
\end{equation}
where $\mathcal{W}$ is a shorthand for the weight function $\mathcal{W}(t') = \{w(t') \mid 0 \leq w(t') \leq 1$ and $\sum_{t_i \leq t' \leq t_j} w(t') = 1\}$. Let $\mathbf{z}^\ast_{t_i: t_j}$ denote a reference trajectory, e.g., $\mathbf{z}^\ast_{t_i: t_j} = (-0.5, \cdots, -0.5)$ corresponding to $50\%$ reduction in total distance traveled from $t_i$ to $t_j$. For each treatment dose trajectory $\mathbf{z}_{t_i: t_j}$, we define its ``cumulative dose" as the weighted difference between $\mathbf{z}_{t_i: t_j}$ and $\mathbf{z}^\ast_{t_i: t_j}$.

\begin{definition}[Cumulative Dose]\rm
\label{def: CD}
Let $\mathbf{z}_{t_i: t_j}$ be a realization of the treatment dose trajectory $\mathbf{Z}_{t_i: t_j}$. Its cumulative dose with respect to the reference trajectory $\mathbf{z}^\ast_{t_i: t_j}$ and the weight function $\mathcal{W}$ is 
\[
\text{CD}(\mathbf{z}_{t_i: t_j}; \mathbf{z}^\ast_{t_i: t_j}, \mathcal{W}) =
\left\|\mathbf{z}_{t_i: t_j} - \mathbf{z}^\ast_{t_i: t_j}\right\|_{\mathcal{W}},
\]
where $\|\cdot\|_{\mathcal{W}}$ is defined in \eqref{eqn: weighted l1 norm}.
\end{definition}

\begin{remark}\rm
\label{remark: cumulative dose}
The cumulative dose of a treatment dose trajectory is defined with respect to a reference trajectory and a weight function. The choices of the reference trajectory and weight function should be guided by expert knowledge so that the cumulative dose reflects some scientifically meaningful aspect of the treatment dose trajectory. For instance, in a longitudinal study of the effect of zidovudine (AZT), an antiretroviral medication, on mortality, \citet{robins2000} defined the cumulative dose to be the aggregate AZT dose during the treatment period, i.e., the reference dose $\mathbf{z}^\ast_{t_0: t} = (0, \cdots, 0)$ and $\text{CD}(\mathbf{z}_{t_0: t}; \mathbf{z}^\ast_{t_0: t}, \mathcal{W}) = \sum_{t_0\leq t' \leq t} z_{t'}$.
\end{remark}

A collection of treatment dose trajectories is said to be ``$\mathcal{W}$-equivalent" if they have the same cumulative dose with respect to the same weight function and reference trajectory. 

\begin{definition}[$\mathcal{W}$-Equivalence]
\label{def: w-equivalence}\rm
Two treatment dose trajectories $\mathbf{z}_{t_i: t_j}$ and $\mathbf{z}'_{t_i: t_j}$ are said to be $\mathcal{W}$-equivalent w.r.t.to the reference trajectory $\mathbf{z}^\ast_{t_i: t_j}$, written as $\mathbf{z}_{t_i: t_j} \overset{\mathcal{W}}{\equiv} \mathbf{z}'_{t_i: t_j}$, if $\text{CD}(\mathbf{z}_{t_i: t_j}; \mathbf{z}^\ast_{t_i: t_j}, \mathcal{W})$ = $\text{CD}(\mathbf{z}'_{t_i: t_j}; \mathbf{z}^\ast_{t_i: t_j}, \mathcal{W})$. Treatment dose trajectories that are equivalent to $\mathbf{z}_{t_i: t_j}$ form an equivalence class and is denoted as 
\[
[\mathbf{z}_{t_i: t_j}]_{\mathcal{W}} = \left\{\mathbf{z}'_{t_i: t_j} \mid \text{CD}(\mathbf{z}'_{t_i: t_j}; \mathbf{z}^\ast_{t_i: t_j}, \mathcal{W}) = \text{CD}(\mathbf{z}_{t_i: t_j}; \mathbf{z}^\ast_{t_i: t_j}, \mathcal{W})\right\}.
\]
\end{definition}

Equipped with Definition \ref{def: CD} and \ref{def: w-equivalence}, we are ready to state a major assumption that facilitates extending a dose-response relationship to longitudinal settings. 

\begin{assumption}[Potential outcomes under $\mathcal{W}$-equivalence]\rm
\label{assump: potential outcome same under CD}
Let $[\mathbf{z}_{t_0: t}]_{\mathcal{W}}$ be an equivalence class as defined in Definition \ref{def: w-equivalence} with respect to $\|\cdot\|_{\mathcal{W}}$ and a reference trajectory $\mathbf{z}^\ast_{t_0: t}$. Then unit-level potential outcomes at time $t$, $Y_{n, t}(\cdot)$, satisfies:
\begin{equation*}
    Y_{n, t}(\mathbf{z}_{t_0: t}) = Y_{n, t}(\mathbf{z}'_{t_0: t}),~\forall~\mathbf{z}_{t_0: t}, \mathbf{z}'_{t_0: t} \in [\mathbf{z}_{t_0: t}]_{\mathcal{W}}.
\end{equation*}
\end{assumption}

\begin{example}\rm
In the study of AZT's effect on mortality, $\mathbf{Z}_{t_0: t}$ represents the AZT dose trajectory from $t_0$ to $t$. Let $Y_{n, 30} = 1$ if unit $n$ dies at time $t = 30$ and $0$ otherwise. Assumption \ref{assump: potential outcome same under CD} applied to $Y_{n, 30}$ then states that patient $n$'s 30-day mortality status depends on the AZT trajectory from $t_0$ to $t$ only through some ``cumulative dose" captured by $\text{CD}(\mathbf{z}_{t_0: t}; \mathbf{z}^\ast_{t_0: t}, \mathcal{W})$ (e.g., the aggregate dose; see Remark \ref{remark: cumulative dose}).
\end{example}

\begin{remark}\rm
Although Assumption \ref{assump: potential outcome same under CD} and its variants are often assumed in the literature on longitudinal studies to reduce the number of potential outcomes (\citealp[Section 7]{robins2000}), its validity needs to be evaluated on a case-by-case basis. We evaluated Assumption \ref{assump: potential outcome same under CD} in the infectious disease dynamics context using standard compartment model before invoking it in our application; see Supplementary Material H for details.
\end{remark}

We now extend the dose-response relationship to a longitudinal setting. 

\begin{definition}[Unit-Level Dose-Response Relationship in Longitudinal Studies]\rm
Let $\text{CD}(\mathbf{z}_{t_0: t}; \mathbf{z}^\ast_{t_0: t}, \mathcal{W})$ be a cumulative dose defined in Definition \ref{def: CD} and $f_n(\cdot; \boldsymbol{\theta}_n)$ a univariate dose-response model parametrized by $\boldsymbol{\theta_n}$ such that $f_n(0; \boldsymbol{\theta_n}) = 0$. Suppose that Assumption \ref{assump: potential outcome same under CD} holds. A unit-level dose-response relationship for unit $n$ states that
\begin{equation}
    \label{eqn: dose-response in longitudinal studies}
    Y_{n, t}(\mathbf{z}_{t_0: t}) - Y_{n, t}(\mathbf{z}^\ast_{t_0: t}) = f_n\left\{\text{CD}(\mathbf{z}_{t_0: t}; \mathbf{z}^\ast_{t_0: t}, \mathcal{W}); \boldsymbol\theta_n\right\}.
\end{equation}
\end{definition}

\begin{remark}\rm
Observe that when $\mathbf{z}_{t_0: t} = \mathbf{z}^\ast_{t_0: t}$, the LHS of \eqref{eqn: dose-response in longitudinal studies} evaluates to $0$ and the RHS evaluates to $f_n\left\{ \text{CD}(\mathbf{z}^\ast_{t_0: t}; \mathbf{z}^\ast_{t_0: t}, \mathcal{W}); \boldsymbol\theta_n\right\} = f_n\left\{0; \boldsymbol\theta_n\right\} = 0$.
\end{remark}

\begin{remark}\rm
Let $\mathbf{z}_{t_0: t}$ and $\mathbf{z}'_{t_0: t}$ be two treatment dose trajectories such that $\mathbf{z}_{t_0: t} \neq \mathbf{z}'_{t_0: t}$ but $\text{CD}(\mathbf{z}_{t_0: t};\mathbf{z}^\ast_{t_0: t}, \mathcal{W}) = \text{CD}(\mathbf{z}'_{t_0: t}; \mathbf{z}^\ast_{t_0: t}, \mathcal{W})$. For the dose-response relationship \eqref{eqn: dose-response in longitudinal studies} to be well-defined, we necessarily have $Y_{n, t}(\mathbf{z}_{t_0: t}) = Y_{n, t}(\mathbf{z}'_{t_0: t})$, which is guaranteed by Assumption \ref{assump: potential outcome same under CD}.
\end{remark}

\begin{remark}\rm
Similar to the static setting considered in Section \ref{sec: setup nbpmatching and potential outcome}, the dose-response relationship \eqref{eqn: dose-response in longitudinal studies} can be thought of as a parsimonious summary of unit-level causal effects from a sustained period of treatment.
\end{remark}

\subsection{Embedding longitudinal data into an experiment and testing a dose-response relationship}
\label{subsec: embed longitudinal data into experiment}
Let $i = 1,2, \cdots, I$ be $I$ pairs of two units matched on the covariate process $\overline{L}_{i1, t} = \overline{L}_{i2, t}$ but $\boldsymbol Z^{\text{obs}}_{i1; t_0: t} \neq \boldsymbol Z^{\text{obs}}_{i2; t_0: t}$. Units $i1$ and $i2$ are each associated with the following two potential outcomes at time $t$:
\begin{equation*}
    Y_{ij, t}(\boldsymbol Z^{\text{obs}}_{i1; t_0: t})~~~\text{and}~~~Y_{ij, t}(\boldsymbol Z^{\text{obs}}_{i2; t_0: t}),~i = 1, \cdots, I,~j = 1, 2,
\end{equation*}
in parallel with Definition \ref{def: define potential outcomes} in the static setting. Write $\mathcal{F}_t = \{\overline{L}_{ij, t}, Y_{ij, t}(\boldsymbol Z^{\text{obs}}_{i1; t_0: t}), Y_{ij, t}(\boldsymbol Z^{\text{obs}}_{i2; t_0: t})$, $i = 1, \cdots, I,~j = 1, 2\}$. Let $ij'$ denote the unit with the minimum cumulative dose in matched pair $i$ and $ij''$ the other unit, and write $\mathbf{Z}^{\text{obs}}_{\wedge; t_0: t}= \{\mathbf{Z}^{\text{obs}}_{1j'; t_0: t}, \cdots, \mathbf{Z}^{\text{obs}}_{Ij'; t_0: t}\}$, and $\mathbf{Z}^{\text{obs}}_{\vee; t_0: t}= \{\mathbf{Z}^{\text{obs}}_{1j''; t_0: t}, \cdots, \mathbf{Z}^{\text{obs}}_{Ij''; t_0: t}\}$. By iteratively applying the sequential randomization assumption, it is shown in the Supplementary Material E that 
\begin{equation}
\label{eqn: cond prob after matching longitudinal}
\begin{split}
    \pi_{i1} =& P(\mathbf{Z}_{i1; t_0:t} = \mathbf{Z}^{\text{obs}}_{i1; t_0:t}, \mathbf{Z}_{i2; t_0:t} = \mathbf{Z}^{\text{obs}}_{i2; t_0:t} \mid \mathcal{F}_t, \mathbf{Z}^{\text{obs}}_{\wedge; t_0:t}, \mathbf{Z}^{\text{obs}}_{\vee; t_0:t}) \\
    =&P(\mathbf{Z}_{i1; t_0:t} = \mathbf{Z}^{\text{obs}}_{i2; t_0:t}, \mathbf{Z}_{i2; t_0:t} = \mathbf{Z}^{\text{obs}}_{i1; t_0:t} \mid \mathcal{F}_t, \mathbf{Z}^{\text{obs}}_{\wedge; t_0:t}, \mathbf{Z}^{\text{obs}}_{\vee; t_0:t}) = \pi_{i2} = 1/2.
\end{split}
\end{equation}

\begin{remark}\rm
In the static setting, it suffices to match on observed covariates to embed data into an approximate experiment; in the longitudinal setting, one needs to match on the covariate process $\overline{L}_t$ including the time-independent covariates and observed outcomes during the treatment period.
\end{remark}

\begin{remark}\rm
Our framework is different from the balance risk set matching of \citet{li2001balanced}. According to \citet{li2001balanced}'s setup, units receive a binary treatment at most once in the entire study period. Our framework is also different from \citet{imai2018matching}. \citet{imai2018matching}'s primary interest is the treatment effect of an intervention at a particular time point $t$; hence, \citet{imai2018matching} pair a subject receiving treatment at time $t$ to subjects with the same treatment dose and covariate process up to time $t - 1$ but not receiving the treatment at time $t$. In sharp contrast, we are focusing on the causal effect of a sustained period of treatment, similar to the setup in \citet{robins2000}. The entire treatment dose trajectory is the unit to be permuted and our design reflects this aspect.
\end{remark}

Consider testing the following dose-response relationship in a longitudinal study:
\begin{equation*}
\begin{split}
    H^{\text{L}}_0: ~~&Y_{ij, t}(\mathbf{z}_{t_0: t}) - Y_{ij, t}(\mathbf{z}^\ast_{t_0: t}) \\
    = &f\left\{ \text{CD}(\mathbf{z}_{t_0: t}; \mathbf{z}^\ast_{t_0: t}, \mathcal{W}); \boldsymbol\theta\right\},\quad\forall i = 1, \cdots, I,~j = 1,2,\quad \text{for some}~\boldsymbol\theta,
\end{split}
\end{equation*}
where $\mathbf{z}^\ast_{t_0: t}$ is a reference trajectory, $\text{CD}(\mathbf{z}_{t_0: t}; \mathbf{z}^\ast_{t_0: t}, \mathcal{W})$ a cumulative dose, and $f(\cdot; \boldsymbol\theta)$ a dose-response relationship of scientific interest. Within each matched pair are two observed treatment dose trajectories $\mathbf{Z}^{\text{obs}}_{i1; t_0:t}$ and $\mathbf{Z}^{\text{obs}}_{i2; t_0:t}$. We observe the potential outcome that $i1$ exhibits under $\mathbf{Z}^{\text{obs}}_{i1; t_0:t}$, i.e., $Y_{i1, t}(\mathbf{Z}^{\text{obs}}_{i1; t_0:t}) = Y^{\text{obs}}_{i1, t}$; moreover, we are able to impute $Y_{i1, t}(\cdot)$ evaluated at $\mathbf{Z}^{\text{obs}}_{i2; t_0:t}$, under $H^{\text{L}}_0$ and Assumption \ref{assump: potential outcome same under CD}:
\begin{equation}
\label{eqn: impute i1 under Z_i2 longitudinal}
    Y_{i1, t}(\mathbf{Z}^{\text{obs}}_{i2; t_0:t}) = Y^{\text{obs}}_{i1, t} + f\left\{ \text{CD}(\mathbf{Z}^{\text{obs}}_{i2; t_0: t}; \mathbf{z}^\ast_{t_0: t}, \mathcal{W}); \boldsymbol\theta\right\} - f\left\{ \text{CD}(\mathbf{Z}^{\text{obs}}_{i1; t_0: t}; \mathbf{z}^\ast_{t_0: t}, \mathcal{W}); \boldsymbol\theta\right\}.
\end{equation}
Similarly, we have $Y_{i2, t}(\mathbf{Z}^{\text{obs}}_{i2; t_0:t}) = Y^{\text{obs}}_{i2, t}$ and can impute:
\begin{equation}
\label{eqn: impute i2 under Z_i1 longitudinal}
    Y_{i2, t}(\mathbf{Z}^{\text{obs}}_{i1; t_0:t}) = Y^{\text{obs}}_{i2, t} + f\left\{ \text{CD}(\mathbf{Z}^{\text{obs}}_{i1; t_0: t}; \mathbf{z}^\ast_{t_0: t}, \mathcal{W}); \boldsymbol\theta\right\} - f\left\{ \text{CD}(\mathbf{Z}^{\text{obs}}_{i2; t_0: t}; \mathbf{z}^\ast_{t_0: t}, \mathcal{W}); \boldsymbol\theta\right\}.
\end{equation}
Table \ref{tbl: science table impute longitudinal} summarizes the observed and imputed information. The problem has now been reduced to the static setting, except that instead of permuting the two scalar treatment doses, we now permute two treatment dose trajectories. Randomization-based testing procedure like the one discussed in Section \ref{sec: randomization inference time-indep} in a static setting can be readily applied to testing (1) $\boldsymbol\theta = \boldsymbol\theta_0$ for a fixed $\boldsymbol\theta_0$ and (2) the validity of a postulated dose-response relationship $H_0^{\text{L}}$.

\begin{table}[h]
\centering
\caption{\small Imputed science table when testing a dose-response relationship in a longitudinal setting. For each unit, one and only one potential outcome is observed; however, the other potential outcome can be imputed under Assumption \ref{assump: potential outcome same under CD} and $H_0^{\text{L}}$.}
\label{tbl: science table impute longitudinal}
\resizebox{\textwidth}{!}{
\fbox{%
\begin{tabular}{cccccccc} \\[-0.8em]
 & & &\multicolumn{2}{c}{Observe One Potential Outcome}
&\multicolumn{2}{c}{Imputed Potential Outcomes} \\ \\[-0.9em]  \cline{4-7} \cline{4-7}\\[-0.9em]
\multirow{3}{*}{\begin{tabular}{c}Units\end{tabular}}  
&\multirow{3}{*}{\begin{tabular}{c}Obs. Treatment\\Dose Trajectory\\ $\mathbf{Z}^{\text{obs}}_{ij; t_0:t}$\end{tabular}}
&\multirow{3}{*}{\begin{tabular}{c}Cumulative\\Dose\\ $\mathbf{Z}^{\text{obs}}_{ij; t_0:t}$\end{tabular}}
&\multirow{3}{*}{\begin{tabular}{c}$Y_{ij, t}(\mathbf{Z}^{\text{obs}}_{i1; t_0: t})$\end{tabular}}
&\multirow{3}{*}{\begin{tabular}{c}$Y_{ij, t}(\mathbf{Z}^{\text{obs}}_{i2; t_0: t})$\end{tabular}}
&\multirow{3}{*}{\begin{tabular}{c}$Y_{ij, t}(\mathbf{Z}^{\text{obs}}_{i1; t_0: t})$\end{tabular}}
&\multirow{3}{*}{\begin{tabular}{c}$Y_{ij, t}(\mathbf{Z}^{\text{obs}}_{i2; t_0: t})$\end{tabular}}
\\[-0.9em] \\ \\ \\[-0.9em]  \\ \hline \\[-0.9em]

\multirow{2}{*}{\begin{tabular}{c}$11$ \end{tabular}} 
&\multirow{2}{*}{\begin{tabular}{c}$\mathbf{Z}^{\text{obs}}_{11; t_0:t}$ \end{tabular}}
& \multirow{2}{*}{\begin{tabular}{c}$\text{CD}(\mathbf{Z}^{\text{obs}}_{11; t_0: t}; \mathbf{z}^\ast_{t_0: t}, \mathcal{W})$ \end{tabular}} 
& \multirow{2}{*}{\begin{tabular}{c}$Y^{\text{obs}}_{11, t}$ \end{tabular}} 
& \multirow{2}{*}{\begin{tabular}{c}$\boldsymbol{?}$ \end{tabular}} 
&\multirow{2}{*}{\begin{tabular}{c}$Y^{\text{obs}}_{11, t}$ \end{tabular}} 
& \multirow{2}{*}{\begin{tabular}{c}
Impute according \\ to scheme \eqref{eqn: impute i1 under Z_i2 longitudinal}
\end{tabular}} 
\\ \\
\multirow{2}{*}{\begin{tabular}{c}$12$ \end{tabular}} 
&\multirow{2}{*}{\begin{tabular}{c}$\mathbf{Z}^{\text{obs}}_{12; t_0:t}$ \end{tabular}}
& \multirow{2}{*}{\begin{tabular}{c}$\text{CD}(\mathbf{Z}^{\text{obs}}_{12; t_0: t}; \mathbf{z}^\ast_{t_0: t}, \mathcal{W})$ \end{tabular}} 
& \multirow{2}{*}{\begin{tabular}{c}$\boldsymbol{?}$ \end{tabular}} & \multirow{2}{*}{\begin{tabular}{c}$Y^{\text{obs}}_{12, t}$ \end{tabular}} 
& \multirow{2}{*}{\begin{tabular}{c}
Impute according \\ to scheme \eqref{eqn: impute i2 under Z_i1 longitudinal}\end{tabular}}
&\multirow{2}{*}{\begin{tabular}{c}$Y^{\text{obs}}_{12, t}$ \end{tabular}} 
\\ \\

$\vdots$ & $\vdots$ & $\vdots$ & $\vdots$ &$\vdots$ &$\vdots$ &$\vdots$ &$\vdots$\\

\multirow{2}{*}{\begin{tabular}{c}$I1$ \end{tabular}} 
&\multirow{2}{*}{\begin{tabular}{c}$\mathbf{Z}^{\text{obs}}_{I1; t_0:t}$ \end{tabular}}
& \multirow{2}{*}{\begin{tabular}{c}$\text{CD}(\mathbf{Z}^{\text{obs}}_{I1; t_0: t}; \mathbf{z}^\ast_{t_0: t}, \mathcal{W})$ \end{tabular}} 
& \multirow{2}{*}{\begin{tabular}{c}$Y^{\text{obs}}_{I1, t}$ \end{tabular}} 
& \multirow{2}{*}{\begin{tabular}{c}$\boldsymbol{?}$ \end{tabular}} 
&\multirow{2}{*}{\begin{tabular}{c}$Y^{\text{obs}}_{I1, t}$ \end{tabular}} 
& \multirow{2}{*}{\begin{tabular}{c}
Impute according \\ to scheme \eqref{eqn: impute i1 under Z_i2 longitudinal}
\end{tabular}} 
\\ \\

\multirow{2}{*}{\begin{tabular}{c}$I2$ \end{tabular}} 
&\multirow{2}{*}{\begin{tabular}{c}$\mathbf{Z}^{\text{obs}}_{I2; t_0:t}$ \end{tabular}}
& \multirow{2}{*}{\begin{tabular}{c}$\text{CD}(\mathbf{Z}^{\text{obs}}_{I2; t_0: t}; \mathbf{z}^\ast_{t_0: t}, \mathcal{W})$ \end{tabular}} 
& \multirow{2}{*}{\begin{tabular}{c}$\boldsymbol{?}$ \end{tabular}} & \multirow{2}{*}{\begin{tabular}{c}$Y^{\text{obs}}_{I2, t}$\end{tabular}} 
& \multirow{2}{*}{\begin{tabular}{c}
Impute according \\ to scheme \eqref{eqn: impute i2 under Z_i1 longitudinal}\end{tabular}}
&\multirow{2}{*}{\begin{tabular}{c}$Y^{\text{obs}}_{I2, t}$ \end{tabular}} 
\\ \\
\end{tabular}}}
\end{table}

\subsection{Time lag and lag-incorporating weights}
One unique aspect of our application is that there is typically a time lag between social distancing and its effect on public-health-related outcomes. We formalize this in Assumption \ref{assump: lag}.

\begin{assumption}[Time Lag]\rm
\label{assump: lag}
The treatment trajectory is said to have a ``$\ell$-lagged effect" on unit $n$'s potential outcomes at time $t$ if
\begin{equation*}
    Y_{n, t}(z_{t_0}, z_{t_1}, \cdots, z_{t - \ell}, z_{t - \ell + 1}, \cdots, z_{t}) = Y_{n, t}(z_{t_0}, z_{t_1}, \cdots, z_{t - \ell}, z'_{t - \ell + 1}, \cdots, z'_{t}),
\end{equation*}
for all $z_{t_0}, \cdots, z_{t - \ell}$,  $z_{t - \ell + 1}, \cdots, z_t$, and $z'_{t - \ell + 1}, \cdots, z'_t$.
\end{assumption}
In words, Assumption \ref{assump: lag} says that the outcome of interest at time $t$ depends on the entire treatment dose trajectory only up to time $t - \ell$. Assumption \ref{assump: lag} holds in particular when $Y_{n, t}$ measures the number of people succumbing to the COVID-19 at time $t$. Researchers estimated that the time lag between contracting COVID-19 and exhibiting symptoms (i.e., the so-called incubation period) had a median of $5.1$ days and could be as long as $11.5$ days (\citealp{lauer2020incubation}), and the time lag between the onset of the COVID-19 symptoms and death ranged from $2$ to $8$ weeks (\citealp{Testa2020, world2020report}). Therefore, it may be reasonable to believe that the number of COVID-19 related deaths at time $t$ does not depend on social distancing practices $\ell$ days immediately preceding $t$ for some properly chosen $\ell$. 
Assumption \ref{assump: lag} may be further combined with Assumption \ref{assump: potential outcome same under CD} to state that unit $n$'s potential outcomes at time $t$ depend on the entire treatment dose trajectory $\mathbf{z}_{t_0: t}$ only via some cumulative dose from time $t_0$ to $t - \ell$ by defining the cumulative dose with respect to some lag-incorporating weight function $\mathcal{W}_{\text{lag}}$ that assigns $0$ weights to $\ell$ treatment doses immediately preceding time $t$.

\begin{remark}\rm
\label{remark: lag and aggregate outcome}
Suppose that the time lag assumption holds for potential outcomes $Y_{n, t}(\cdot)$, $\cdots$, $Y_{n, t+\ell-1}(\cdot)$, and let $g: \mathbb{R}^l \mapsto \mathbb{R}$ be a function that maps these $\ell$ potential outcomes to an aggregate outcome $g\{Y_{n, t}(\cdot), \cdots, Y_{n, t + \ell - 1}(\cdot)\}$. One immediate consequence of Assumption \ref{assump: lag} is that $g\{Y_{n, t}(\cdot), \cdots, Y_{n, t + \ell - 1}(\cdot)\}$ depends on the entire treatment dose trajectory $\mathbf{Z}_{t_0: t+\ell-1}$ only via $\mathbf{Z}_{t_0: t-1}$; moreover, we may invoke Assumption \ref{assump: potential outcome same under CD} and further state that $g\{Y_{n, t}(\cdot), \cdots, Y_{n, t + \ell - 1}(\cdot)\}$ depends on the entire treatment dose trajectory $\mathbf{Z}_{t_0: t+\ell-1}$ only via some cumulative dose of $\mathbf{Z}_{t_0: t-1}$. Dose-response relationships, statistical matching, and testing procedures described in Section \ref{subsec: CD, W-equivalence, dose-response in longitudinal setting} and \ref{subsec: embed longitudinal data into experiment} then hold by replacing $Y_{n, t}(\cdot)$ with the aggregate outcome $g\{Y_{n, t}(\cdot), \cdots, Y_{n, t + \ell - 1}(\cdot)\}$ where appropriate. Details are provided in Supplementary Material F.
\end{remark}

\subsection{Incorporating interference}
\label{subsec: incorporate interference longitudinal}
One may further allow the outcome of interest of unit $ij$ to depend not only on its own cumulative dose, but also the cumulative doses of neighboring units, as described in Section \ref{sec: interference}. Let $\vec{\mathbf{Z}}_{t_0:t} = (\mathbf{Z}_{11; t_0: t}, \cdots, \mathbf{Z}_{I2; t_0: t})$
denote the random treatment dose trajectories from $t_0$ to $t$ of all study units, $\vec{\mathbf{z}}_{t_0:t}$ its realization, $\vec{\mathbf{z}}^\ast_{t_0:t} = (\mathbf{z}^\ast_{t_0: t}, \cdots, \mathbf{z}^\ast_{t_0: t})$ a collection of reference dose trajectories, and $Y_{ij, t}(\vec{\mathbf{Z}}_{t_0:t})$ the potential outcome. Finally, collect the cumulative doses of all study units in $\vec{\mathbf{z}}_{\text{cumu}} = (\text{CD}(\mathbf{z}_{11; t_0: t}; \mathbf{z}^\ast_{t_0: t}, \mathcal{W}), \cdots, \text{CD}(\mathbf{z}_{I2; t_0: t}; \mathbf{z}^\ast_{t_0: t}, \mathcal{W}))$. We stress that each entry of $\vec{\mathbf{Z}}_{t_0:t}$ is itself a random trajectory, while each entry of $\vec{\mathbf{z}}_{\text{cumu}}$ is a scalar. Synthesizing our development in Section \ref{sec: interference} and \ref{subsec: embed longitudinal data into experiment}, we consider testing a dose-response relationship in a longitudinal study under interference by modeling the contrast between $Y_{ij, t}(\vec{\mathbf{z}}_{t_0:t})$ and $Y_{ij, t}(\vec{\mathbf{z}}^\ast_{t_0:t})$. Combining Principle I and II in Section \ref{sec: interference} with Assumption \ref{assump: potential outcome same under CD}, we have a causal null hypothesis of the form
\begin{equation*}
H^{\text{L}}_{0, \text{interference}}:
    Y_{ij, t}(\vec{\mathbf{z}}_{t_0:t}) - Y_{ij, t}(\vec{\mathbf{z}}^\ast_{t_0:t}) = f\left\{ \text{CD}(\mathbf{z}_{ij; t_0: t}; \mathbf{z}^\ast_{t_0: t}, \mathcal{W}); \boldsymbol\theta\right\} + g(\langle\vec{\mathbf{z}}_{\text{cumu}}, \mathbf{G}_{ij,\bigcdot}\rangle),
\end{equation*}
where $f\left\{ \text{CD}(\mathbf{z}_{ij; t_0: t}; \mathbf{z}^\ast_{t_0: t}, \mathcal{W}); \boldsymbol\theta\right\}$ captures the dose-response direct effect and $g(\langle\vec{\mathbf{z}}_{\text{cumu}}, \mathbf{G}_{ij,\bigcdot}\rangle)$ models a spillover effect that depends only on the cumulative doses of $ij$'s neighboring units. Simple parametric models as described in Section \ref{subsec: dose-response under interference} can be readily applied to model the spillover effect. By imputing under $H^{\text{L}}_{0, \text{interference}}$ and permuting the two treatment dose trajectories within each matched pair as described in Section \ref{subsec: embed longitudinal data into experiment}, one can readily conduct randomization-based inference to construct confidence sets for structural parameters in the dose-response relationship while treating interference parameters in the $g(\cdot)$ model as sensitivity parameters.

\section{Social distancing and COVID-19 during phased reopening: study design}

\subsection{Data: time frame, granularity, cumulative dose, outcome, and covariate history}
\label{subsec: data}
The first state in the U.S. that reopened was Georgia at April 24th, 2020. We hence consider data from April 27th, the first Monday following April 24th, to August 2nd, the first Sunday in August in the primary analysis. We choose a Monday (April 27th) as the baseline period and a Sunday (August 2rd) as the endpoint because social distancing and public-health-related outcomes data exhibited consistent weekly periodicity (\citealp{unnikrishnan2020globally}). 

We analyze the data at a county-level granularity and use the county-level percentage change in the total distance traveled compiled by $\textsf{Unacast}\textsuperscript{\texttrademark}$ as the continuous, time-varying treatment dose.  We consider a two-month treatment period from April 27th (Monday) to June 28th (Sunday). According to the data compiled by $\textsf{Unacast}\textsuperscript{\texttrademark}$, counties cut social mobility by at most $50\%$ during most of the phased reopening; hence, we set the reference dose trajectory to be $-0.5$ throughout the treatment period and define a notion of cumulative dose with respect to this reference dose trajectory and a uniform weighting scheme that assigns equal weight to each day during the treatment period. In a sensitivity analysis, we further repeated all dose-response analyses using different notions of cumulative dose based on different weighting schemes. In the Supplementary Material H, we assess the appropriateness of Assumption \ref{assump: potential outcome same under CD} in the context of standard epidemiological models using simulation studies. The primary outcome of interest is the cumulative COVID-19 related death toll per $100,000$ people from June 29th (Monday) to August 2nd (Sunday), a total of five weeks. The county-level COVID-19 case number and death toll are both obtained from the New York Times COVID-19 data repository (\citealp{NYT_data2020}).

As discussed in Section \ref{subsec: embed longitudinal data into experiment}, we matched counties similar on covariates, including time-independent covariates and time-dependent covariate processes, in order to embed data into an approximate randomized experiment. Specifically, we matched on the following time-independent baseline covariates: female (\%), black (\%), Hispanic (\%), above $65$ (\%), smoking (\%), driving alone to work (\%), flu vaccination (\%), some college (\%), number of membership associations per $10,000$ people, rural ($0/1$), poverty rate (\%), population, and population density (residents per square mile). These county-level covariates data were derived from the census data collected by the United States Census Bureau and the County Health Rankings and Roadmaps Program (\citealp{remington2015county}). Moreover, we matched on the number of new COVID-19 cases and new COVID-19 related deaths per $100,000$ people every week from April 20th - 26th to June 23th - 29th.

\subsection{Statistical matching, matched samples, and assessing balance}
A total of $1,211$ matched pairs of two counties were formed using optimal nonbipartite matching (\citealp{lu2001matching,lu2011optimal}). We matched exactly on the covariate ``rural (0/1)" for later subgroup analysis and balanced all other $32$ covariates. We added a mild penalty on the cumulative dose so that two counties within the same matched pair had a tangible difference in their cumulative doses, and added $20\%$ sinks to eliminate $20\%$ of counties for whom no good match can be found (\citealp{baiocchi2010building, lu2011optimal}). Following the advice in \citet{rubin2007design}, the design was conducted without any access to the outcome data in order to assure the objectivity of the design.

\begin{figure}[ht]
    \centering
    \includegraphics[width = 0.95\columnwidth]{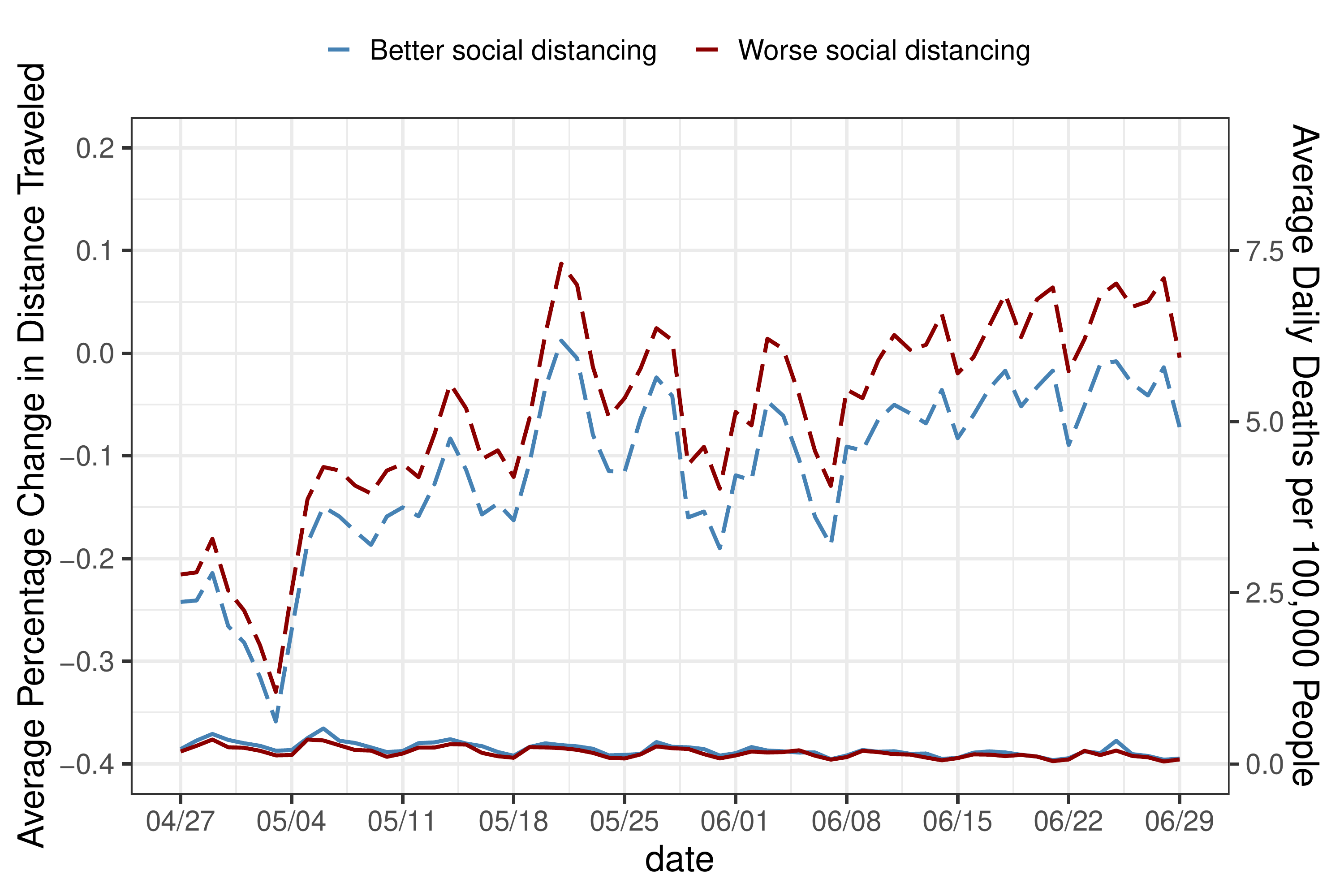}
    \caption{\small Trajectories of the average daily percentage change in total distance traveled (dashed lines) and average daily COVID-19 related death toll per 100,000 people (solid lines) in $1,211$ better social distancing counties (blue) and $1,211$ worse social distancing counties (red). We saw a sharp contrast in the level of social distancing but little difference in the COVID-19 related death during this treatment period.}
    \label{fig: treatment period}
\end{figure}

Within each matched pair, the county with smaller cumulative dose is referred to as the ``better social distancing" county, and the other ``worse social distancing" county. Appendix \ref{appA} shows where the $1,211$ better social distancing counties and the other $1,211$ worse social distancing counties are located in the U.S., and Figure~\ref{fig: treatment period} plots the average daily percentage change in total distance traveled and the average daily COVID-19 related death toll per $100,000$ people during the treatment period (April 27th to June 28th) in two groups. It is evident that two groups differ in their extent of social distancing, but are very similar in their daily COVID-19 related death toll per $100,000$ people during the treatment period. Finally, Appendix \ref{appB} summarizes the balance of all $33$ covariates in two groups after matching. All variables have standardized differences less than $0.15$ and are considered sufficiently balanced (\citealp{rosenbaum2002observational}). In Supplementary Material G.1, we further plot the cumulative distribution functions of important variables in two groups. A detailed pre-analysis plan, including matched samples and specification of a primary analysis and three secondary analyses, can be found via \url{doi: 10.13140/RG.2.2.23724.28800}.

\section{Social distancing and COVID-19 during phased reopening: outcome analysis}
\subsection{Primary analysis: causal null hypothesis regarding the death toll}
\label{subsec: primary analysis}
Fix $t_0 = $ April 27th and $T = $ June 28th. Let $\mathbf{Z}_{t_0: T} = \mathbf{z}_{t_0: T}$ denote a treatment dose trajectory from $t_0$ to $T$ and  $Y_{t}(\cdot)$ the potential COVID-19 related deaths at time $t$. We specify the time-lag parameter $\ell = 35$ so that $T + \ell$ corresponds to August 2nd. As discussed in Remark \ref{remark: lag and aggregate outcome}, we consider the aggregate outcome $Y_{\text{agg}}(\cdot) = g\{Y_{T+1}(\cdot), \cdots, Y_{T + \ell}(\cdot)\} = \sum_{T + 1 \leq t \leq T+\ell} Y_{t} (\cdot)$. Our primary analysis tests the following causal null hypothesis for the $1,211 \times 2 = 2,422$ counties in our matched samples:
\begin{equation*}
    \label{eqn: primary hypothesis}
    H_{0, \text{primary}}:~~Y_{ij, \text{agg}}(\mathbf{z}_{t_0: T}) - Y_{ij, \text{agg}}(\mathbf{z}^\ast_{t_0: T}) = 0,~\forall \mathbf{z}_{t_0: T},~\forall i = 1, \cdots, I = 1211,~j = 1, 2.
\end{equation*}
This null hypothesis states that the treatment dose trajectory from April 27th to June 28th had no effect whatsoever on the COVID-19 related death toll from June 29th to August 2nd.

\begin{figure}[ht]
   \centering
     \subfloat{\includegraphics[width = 0.49\columnwidth]{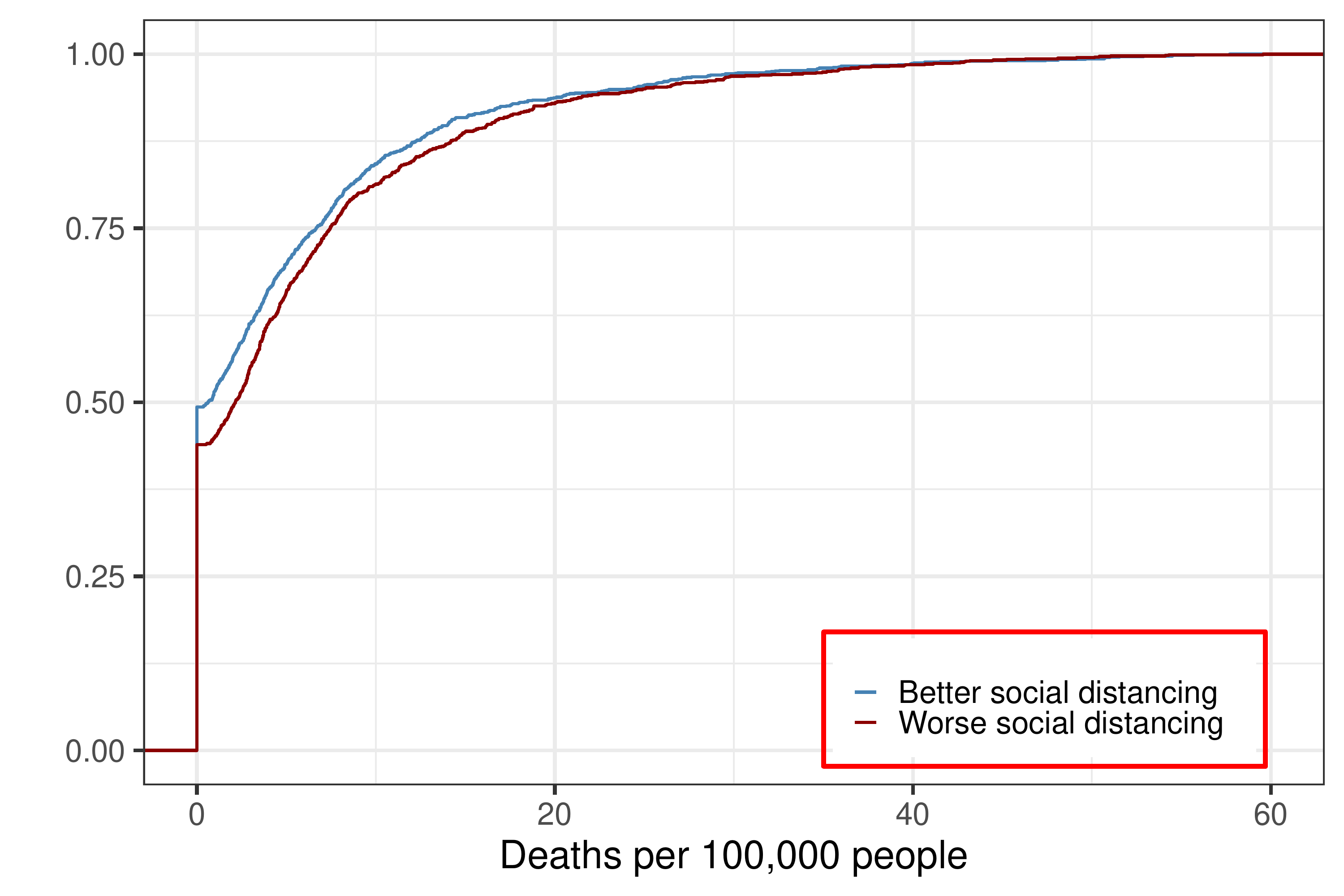}}
     \subfloat{\includegraphics[width = 0.49\columnwidth]{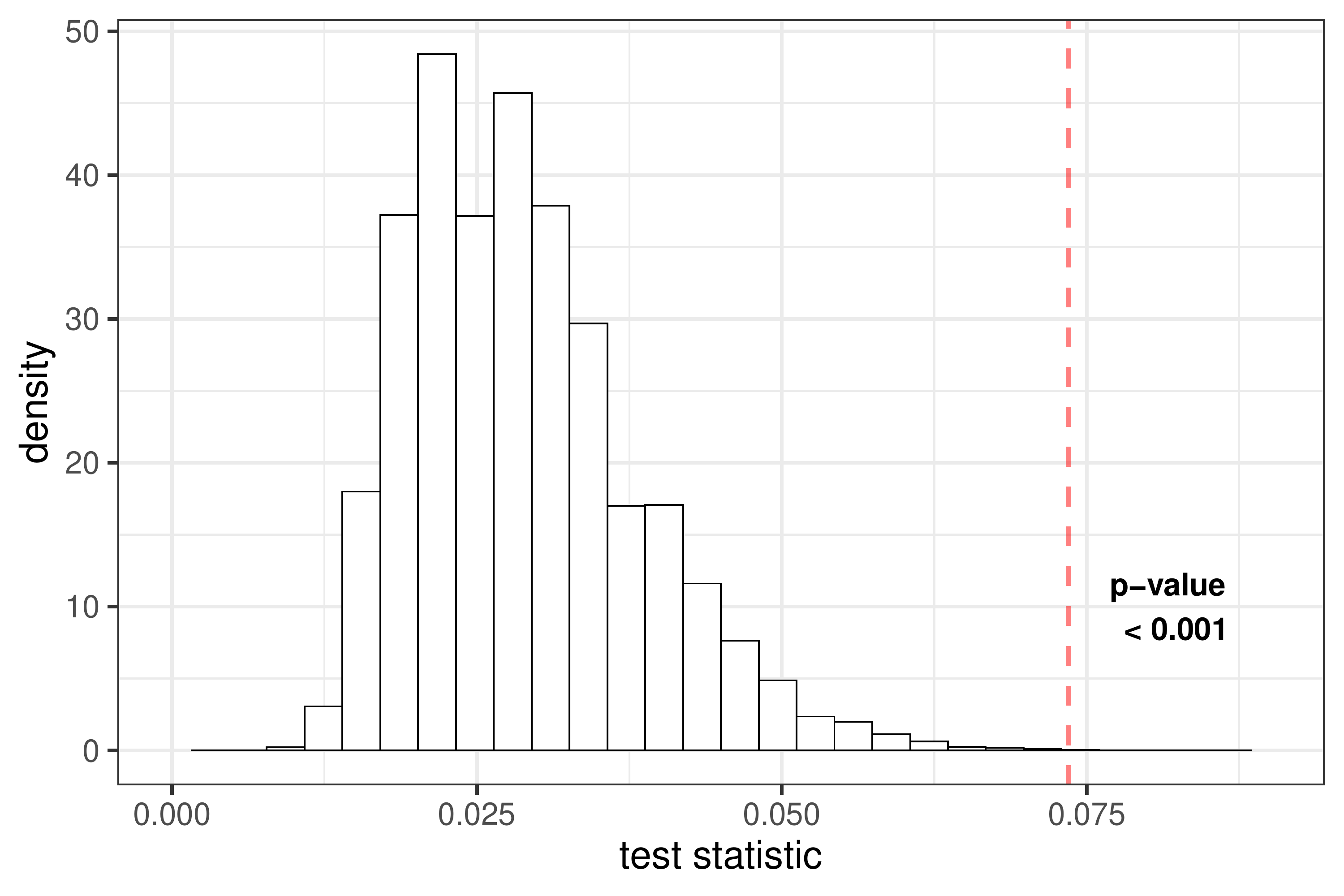}}\\
      \subfloat{\includegraphics[width = 0.49\columnwidth]{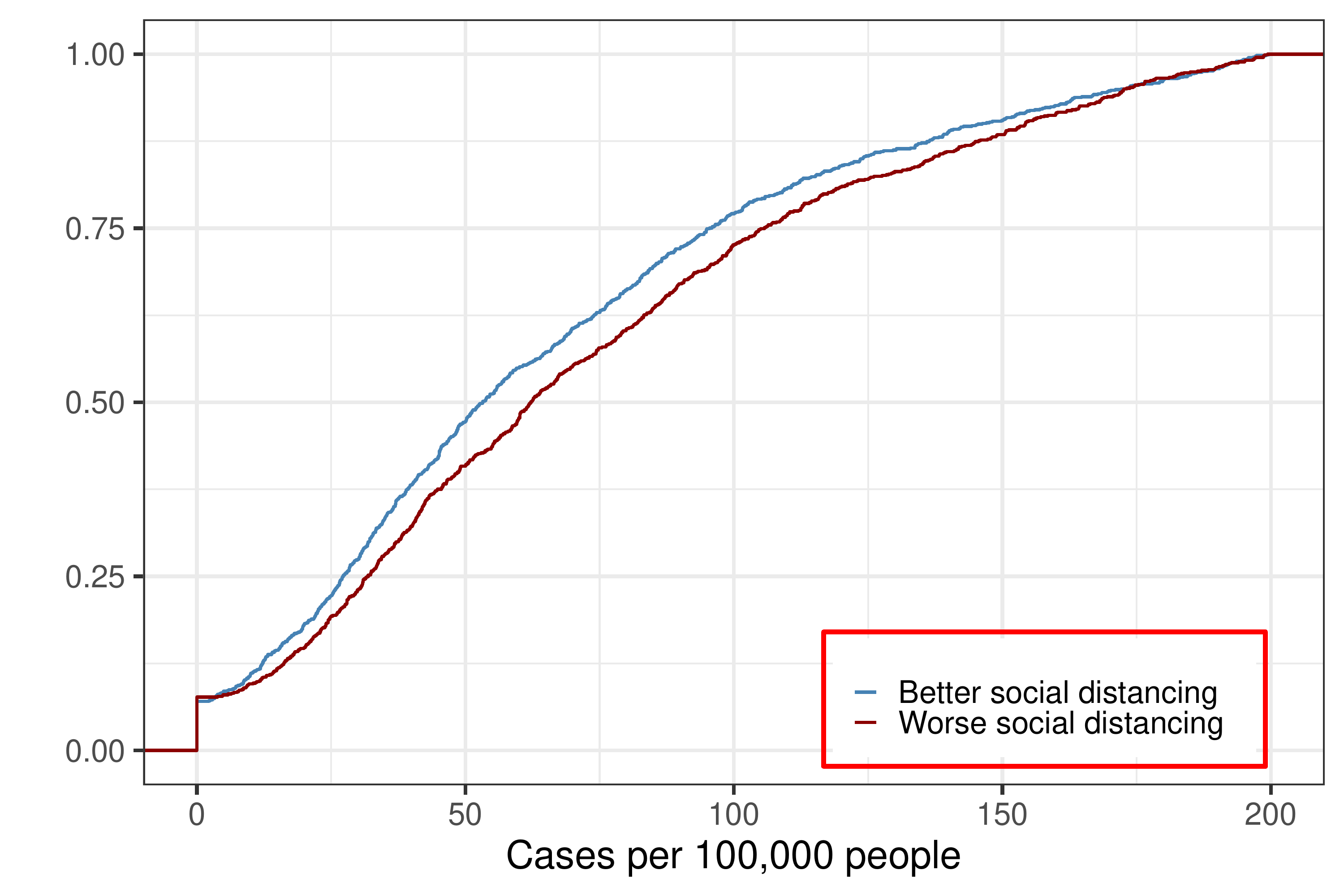}}
     \subfloat{\includegraphics[width = 0.49\columnwidth]{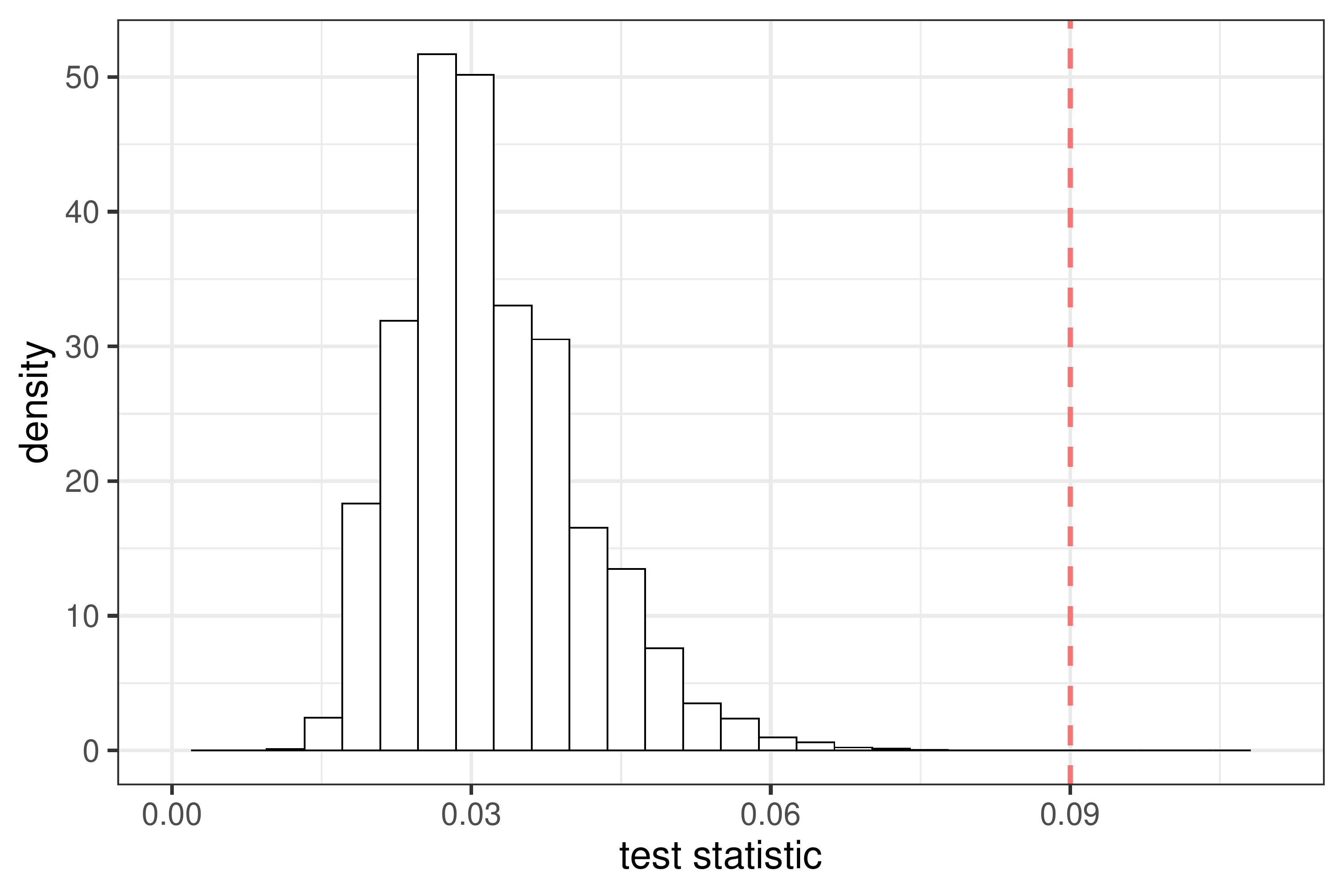}}
     \caption{\small Top left panel: CDFs of cumulative COVID-19 related deaths per $100,000$ people in the better social distancing (blue) and worse social distancing (red) groups. Top right panel: randomization-based reference distribution. The exact p-value is $2.06 \times 10^{-4}$. Bottom left panel: CDFs of cumulative COVID-19 cases per $100,000$ people in the better social distancing (blue) and worse social distancing (red) groups. Bottom right panel: randomization-based reference distribution. The exact p-value is less than $10^{-5}$.}
\label{fig: real data primary analysis and secondary outcome}
\end{figure}

The top left panel of Figure \ref{fig: real data primary analysis and secondary outcome} plots $\widehat{F}_{\text{min}}$ (CDF of the better social distancing counties' observed outcomes) and $\widehat{F}^{\text{tr}}_{\max}$ (CDF of the worse social distancing counties' observed outcomes under $H_{0, \text{primary}}$); we calculate the test statistic $t_{\text{KS}} = 0.735$ and contrast it to a reference distribution generated using $1,000,000$ samples from all possible $2^{1211}$ randomizations; see the top right panel of Figure \ref{fig: real data primary analysis and secondary outcome}. In this way, an exact p-value equal to $2.06 \times 10^{-4}$ is obtained and the causal null hypothesis $H_{0, \text{primary}}$ is rejected at $0.05$ level. Moreover, as detailed in Section \ref{sec: interference} and \ref{subsec: incorporate interference longitudinal}, rejecting the null hypothesis $H_{0, \text{primary}}$ also implies rejecting the null hypothesis of no direct or spillover effect under \emph{arbitrary} interference pattern.

We further conducted two sensitivity analyses to assess the no unmeasured confounding assumption and the time lag assumption we made in the primary analysis. In the first sensitivity analysis, we allowed the dose trajectory assignment probability $\pi_{i1}$ and $\pi_{i2}$ as in \eqref{eqn: cond prob after matching longitudinal} to be biased from the randomization probability and then generated the reference distribution with this biased randomization probability; specifically, we considered a biased treatment assignment model where $\log(\Gamma_i) = \log\{\pi_{i1}/\pi_{i2}\}$ in each matched pair $i$ was proportional to the absolute difference in the cumulative doses of two units in the pair ($\pi_{i1} = \pi_{i2} = 1/2$ and $\Gamma_i = 1$ in a randomized experiment for all $i$). We found that our primary analysis conclusion would hold up to $\Gamma_i$ having a median as large as $3.82$. See Supplementary Material G.3.1 for details. In the second sensitivity analysis, we repeated the primary analysis using a shorter time lag $l = 28$ days and the result was similar; see Supplementary Material G.3.2 for details.

Our primary analysis results suggested that different social distancing trajectories during the treatment period had an effect on the COVID-19-related death toll in the subsequent weeks. This causal conclusion stands under arbitrary interference pattern and is robust to unmeasured confounding.

\subsection{Secondary analysis I: secondary outcome}
\label{subsec: secondary analysis I: secondary outcome}
Let $Y_{ij, \text{case}, \text{agg}}$ denote the cumulative COVID-19 cases per $100,000$ people from June $29$th to July $12$th (corresponding to time lag $l = 14$ days), as specified in our pre-analysis plan. We test the following null hypothesis concerning the secondary outcome $Y_{ij, \text{case}, \text{agg}}$:
\begin{equation*}
    H_{0, \text{secondary}}:~~Y_{ij, \text{case}, \text{agg}}(\mathbf{z}_{t_0: T}) - Y_{ij, \text{case}, \text{agg}}(\mathbf{z}^\ast_{t_0: T}) = 0,~\forall \mathbf{z}_{t_0: T},~\forall i = 1, \cdots, I = 1211,~j = 1, 2.
\end{equation*}
The exact p-value is less than $10^{-5}$; see the bottom panels of Figure \ref{fig: real data primary analysis and secondary outcome}. In a sensitivity analysis, we repeated the test with a shorter time lag $l = 10$ days and the result was similar; see Supplementary Material G.3.3 for details. Our result suggests strong evidence that social distancing from April 27th to June 28th had an effect on cumulative COVID-19 cases per $100,000$ people from June $29$th to July $12$th in our matched samples.

\subsection{Secondary analysis II: explore dose-response relationship}
\label{subsec: secondary analysis II dose-response relationship}
Let $\mathbf{z}^\ast_{t_0: T}$ denote a reference trajectory equal to the $-0.50$ for all $t_0 \leq t \leq T$ (corresponding to $50\%$ reduction in total distance traveled from April 27th to June 28th), $\mathcal{W}_{\text{lag}}$ a weight function that assigns equal weight to all $t$ such that $t_0 \leq t \leq T$ and $0$ otherwise, and a cumulative dose $\text{CD}(\mathbf{z}_{t_0: T}; \mathbf{z}^\ast_{t_0: T}, \mathcal{W}_{\text{lag}})$ defined with respect to $z^\ast_{t_0: T}$ and $\mathcal{W}_{\text{lag}}$. We invoke Assumption \ref{assump: potential outcome same under CD} so that $Y_{\text{agg}}(\cdot)$ depends on $\mathbf{Z}_{t_0: T} = \mathbf{z}_{t_0: T}$ only via $\text{CD}(\mathbf{z}_{t_0: T}; \mathbf{z}^\ast_{t_0: T}, \mathcal{W}_{\text{lag}})$, and consider testing the dose-response kink model concerning the aggregate case number $Y_{ij, \text{case}, \text{agg}}$:
\begin{equation}
\label{eqn: Real data test kink model}
\begin{split}
    H_{0, \text{kink}}:~~&Y_{ij, \text{case}, \text{agg}}(\mathbf{z}_{t_0: T}) - Y_{ij, \text{case}, \text{agg}}(\mathbf{z}^\ast_{t_0:T}) = 0,~\forall \mathbf{z}_{t_0: T} ~\text{such that}~ \text{CD}(\mathbf{z}_{t_0: T}; \mathbf{z}^\ast_{t_0: T}, \mathcal{W}_{\text{lag}}) \leq \tau, \\ & \text{and}~~\log\{Y_{ij,\text{case}, \text{agg}}(\mathbf{z}_{t_0: T})\} - \log\{Y_{ij, \text{case}, \text{agg}}(\mathbf{z}^\ast_{t_0:T})\} = \beta\cdot\{\text{CD}(\mathbf{z}_{t_0: T}; \mathbf{z}^\ast_{t_0: T}, \mathcal{W}_{\text{lag}}) - \tau\}, \\
    &\hspace{1 cm}\forall\mathbf{z}_{t_0: T}~\text{such that}~\text{CD}(\mathbf{z}_{t_0: T}; \mathbf{z}^\ast_{t_0: T}, \mathcal{W}_{\text{lag}}) > \tau,~~\forall i = 1, \cdots, I = 1211, j = 1, 2.
\end{split}
\end{equation}
This dose-response relationship states that the potential COVID-19 cases from June 29th to July 12th remains the same as the potential outcome under $\mathbf{Z}_{t_0: T} = \mathbf{z}^\ast_{t_0: T}$, i.e., strict social distancing that reduces total distance traveled by $50\%$ everyday from April 27th to June 28th, when the cumulative dose (defined w.r.t. $\mathbf{z}^\ast_{t_0: T}$ and $\mathcal{W}_{\text{lag}}$) is less than some threshold $\tau$; after the cumulative dose exceeds this threshold, the COVID-19 case number increases exponentially at a rate proportional to how much the cumulative dose exceeds the threshold. 

\begin{figure}[h]
    \centering
    \subfloat{\includegraphics[width = 0.49\columnwidth]{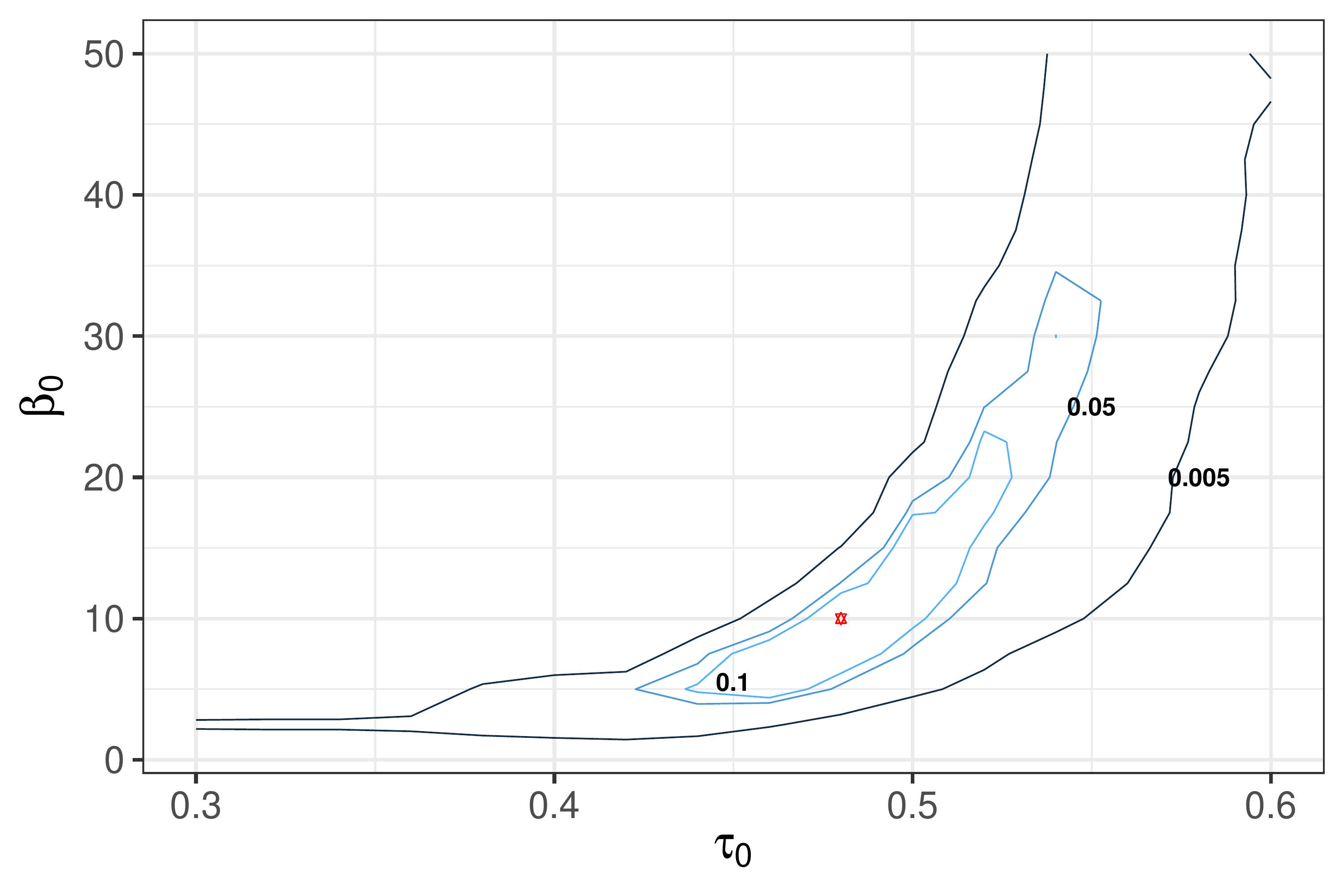}}
     \subfloat{\includegraphics[width = 0.49\columnwidth]{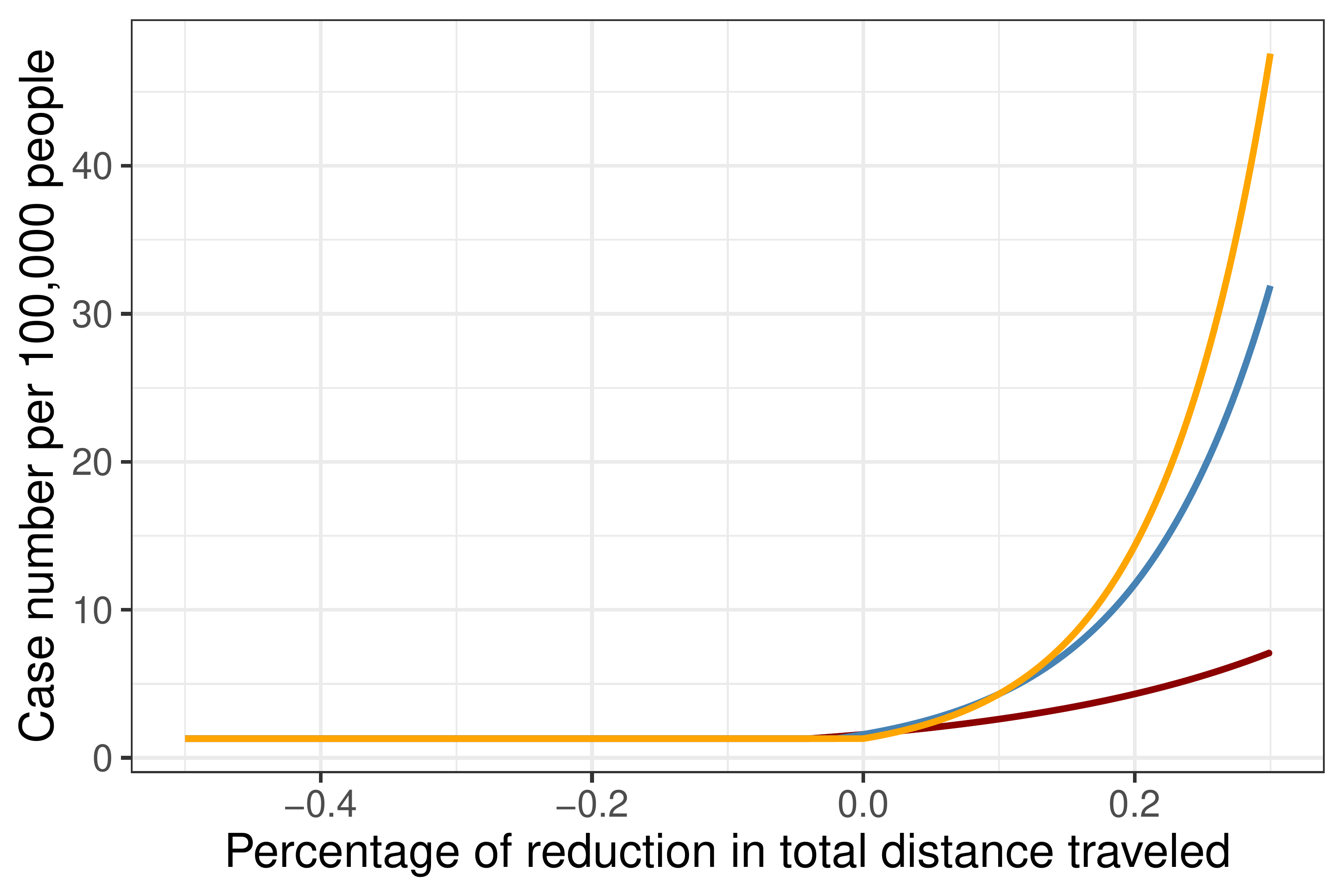}}
    \caption{\small Left panel: contour plot of p-values when testing $H_{0, \text{kink}}$ as in \eqref{eqn: Real data test kink model} against $\tau = \tau_0$ and $\beta = \beta_0$. Maximum p-value is obtained at $(\tau_0, \beta_0) = (0.48, 10.0)$ (red marker). Three isopleths ($0.1$, $0.05$, and $0.005$) are plotted. Right panel: dose-response relationships for selected $(\tau_0, \beta_0)$ in the $0.1$ confidence set with baseline $Y_{ij, \text{case}, \text{agg}}(\mathbf{z}^\ast_{t_0: T})$ equal to $1$ per $100,000$. The red line corresponds to $(\tau_0, \beta_0) = (0.46, 5.0)$, blue line $(\tau_0, \beta_0) = (0.48, 10.0)$, and orange line $(\tau_0, \beta_0) = (0.50, 12.0)$.}
    \label{fig: real data secondary analysis contour and dose-response curves}
\end{figure}

We tested \eqref{eqn: Real data test kink model} for different $\tau = \tau_0$ and $\beta = \beta_0$ combinations; the maximum p-value obtained at $(\tau_0, \beta_0) = (0.48, 10.0)$ is equal to $0.417$ and hence the kink model \eqref{eqn: Real data test kink model} cannot be rejected. The left panel of Figure \ref{fig: real data secondary analysis contour and dose-response curves} plots the level-$0.1$ and level-$0.05$ confidence sets of $(\tau, \beta)$. The right panel of Figure \ref{fig: real data secondary analysis contour and dose-response curves} plots three dose-response curves with baseline (i.e., $50\%$ reduction) case number equal to $1$ per $100,000$ people for three selected $(\tau_0, \beta_0)$ pairs in the level-$0.1$ confidence set. In a sensitivity analysis, we repeated the analysis by considering two different specifications of cumulative dose, one assigning more weight to early days of the phased reopening and the other towards the end of the phased reopening. Confidence sets results look similar under different specifications; see Supplementary Material G.3.4 for details.

The confidence set of the threshold parameter $\tau$ is tightly centered around $0.5$, suggesting that as a county's average percentage change in total distance traveled from April 27th to June 28th increases from $-50\%$ to around $-5\%$ to $5\%$, the potential COVID-19 cases number from June 29th to July 12th would largely remain unchanged; however, once beyond this threshold, the case number would rise exponentially and could incur an increase as large as $10$-fold when a county's average distance traveled increased by about $20\%$ compared to the pre-coronavirus level.

\subsection{Secondary analysis III: subgroup analysis and differential dose-response relationship}
\label{subsec: additional secondary analyses}
\begin{figure}[h]
    \centering
    \subfloat{\includegraphics[width = 0.49\columnwidth]{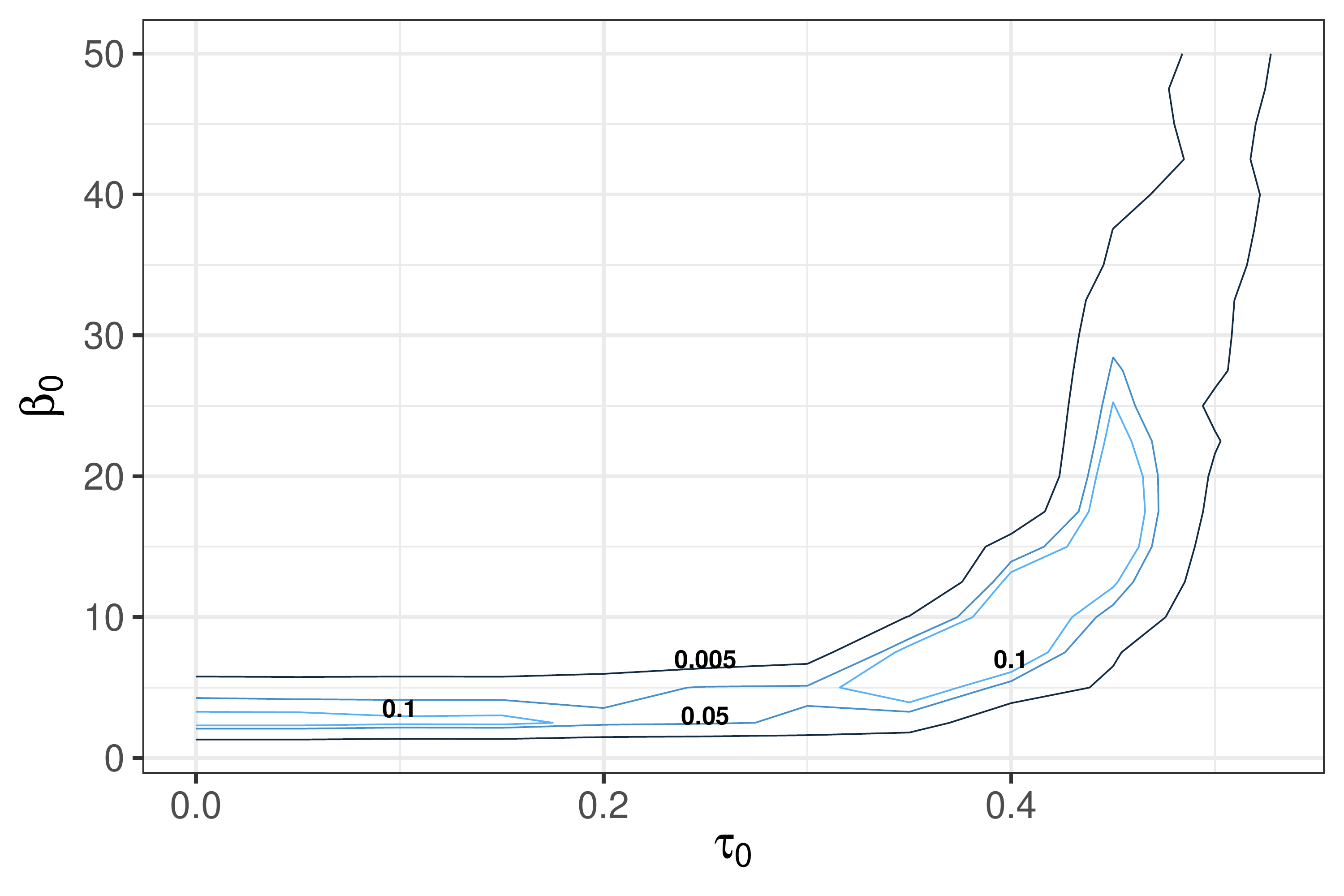}}
     \subfloat{\includegraphics[width = 0.49\columnwidth]{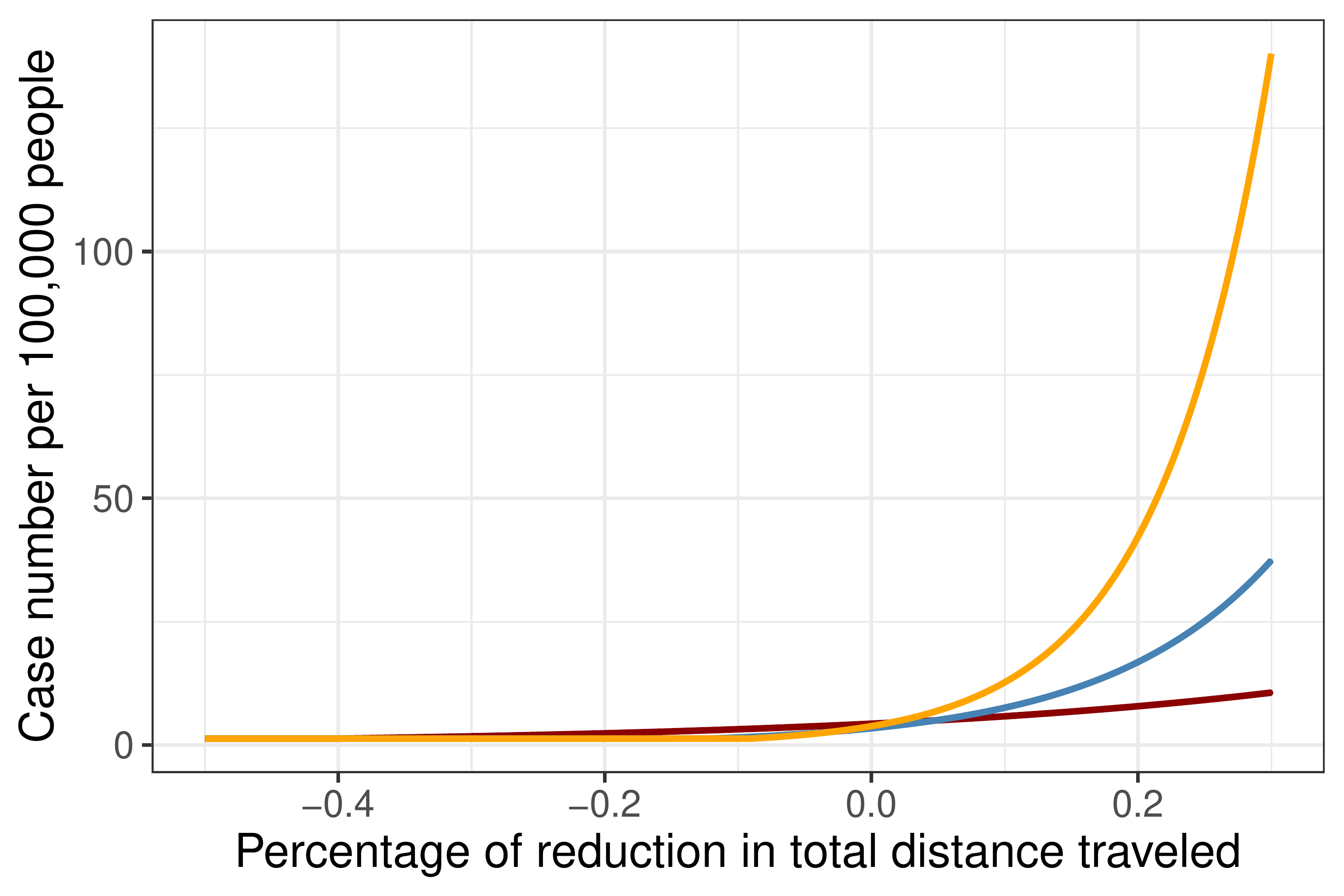}}\\
     \subfloat{\includegraphics[width = 0.49\columnwidth]{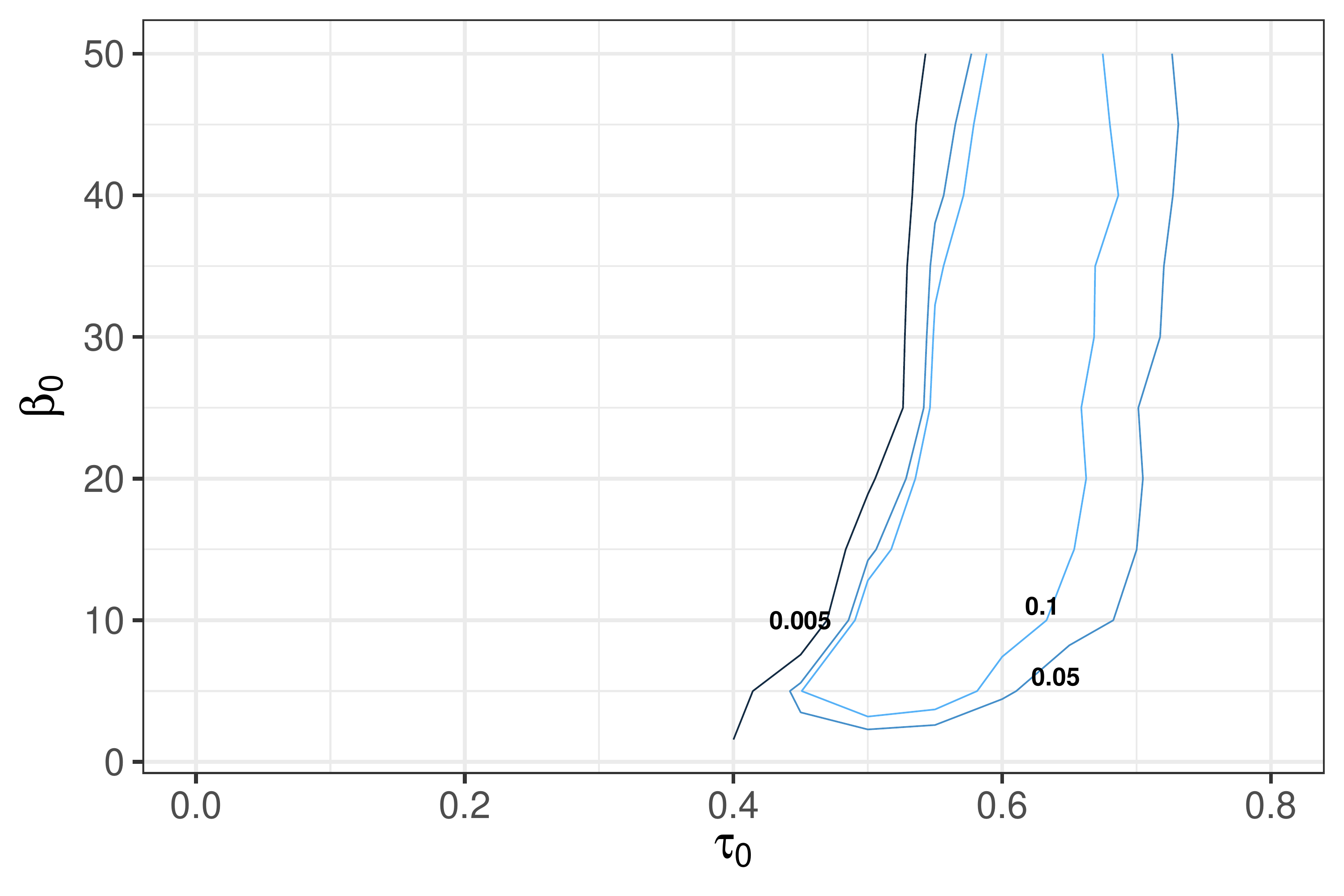}}
     \subfloat{\includegraphics[width = 0.49\columnwidth]{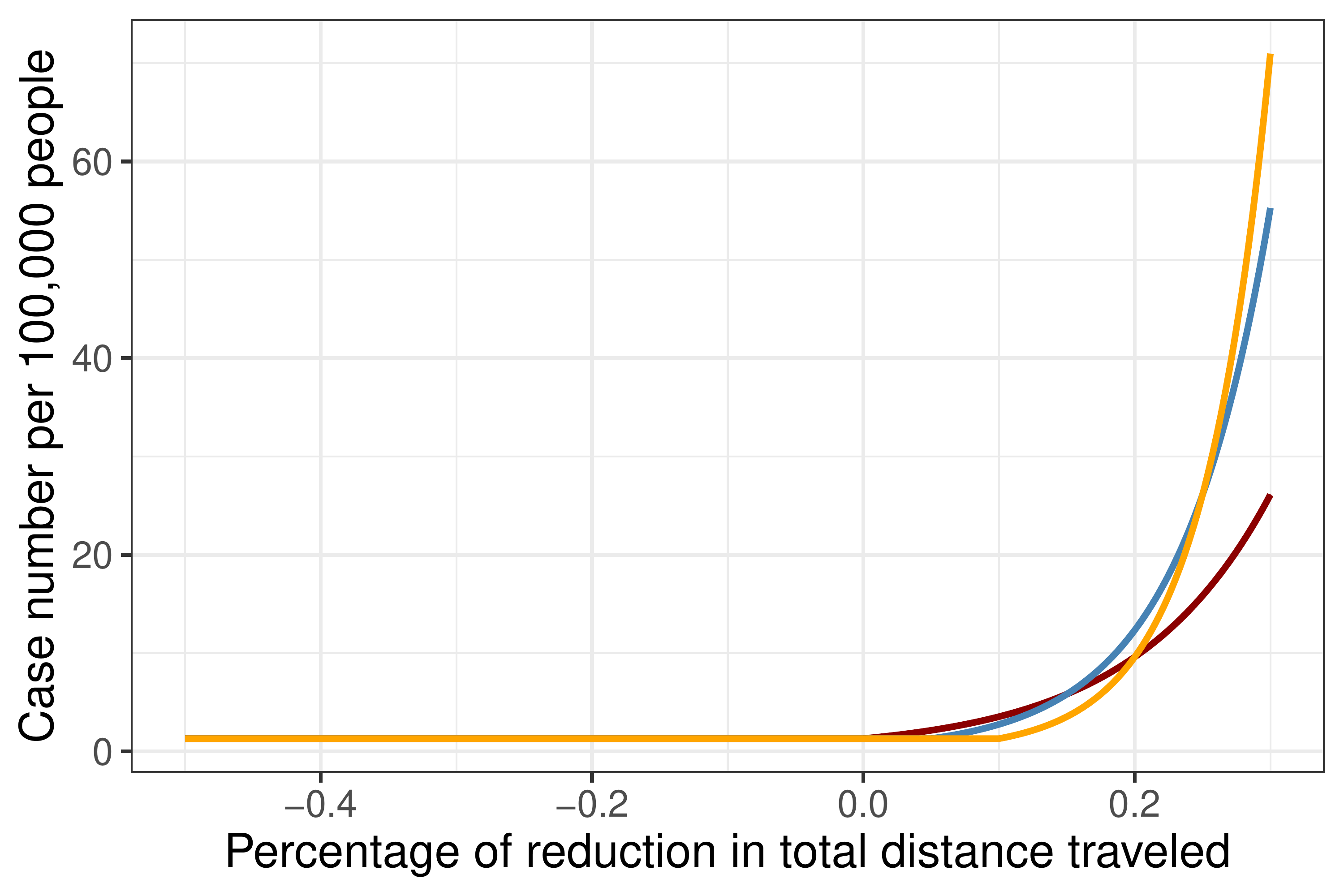}}
    \caption{\small Top left panel: contour plot of p-values when testing $H_{0, \text{kink}}$ against $\tau = \tau_0$ and $\beta = \beta_0$ in $462$ matched pairs of $2$ non-rural counties. Three isopleths ($0.1$, $0.05$, and $0.005$) are plotted. Top right panel: dose-response relationships for selected $(\tau_0, \beta_0)$ in the $0.1$ confidence set as in the top left panel with baseline $Y_{ij, \text{case}, \text{agg}}(\mathbf{z}^\ast_{t_0: T})$ equal to $1$ per $100,000$. The red line corresponds to $(\tau_0, \beta_0) = (0.10, 3.0)$, blue line $(\tau_0, \beta_0) = (0.38, 8.0)$, and orange line $(\tau_0, \beta_0) = (0.41, 12.0)$. Bottom left panel: contour plot of p-values when testing $H_{0, \text{kink}}$ against $\tau = \tau_0$ and $\beta = \beta_0$ in $749$ matched pairs of rural counties. Three isopleths ($0.1$, $0.05$, and $0.005$) are plotted. Bottom right panel: dose-response relationships for selected $(\tau_0, \beta_0)$ in the $0.1$ confidence set as in the bottom left panel with baseline $Y_{ij, \text{case}, \text{agg}}(\mathbf{z}^\ast_{t_0: T})$ equal to $1$ per $100,000$. The red line corresponds to $(\tau_0, \beta_0) = (0.5, 10.0)$, blue line $(\tau_0, \beta_0) = (0.55, 15.0)$, and orange line $(\tau_0, \beta_0) = (0.60, 20.0)$.}
    \label{fig: real data metro and rural contour and dose-response curves}
\end{figure}

We also conducted subgroup analyses by repeating the primary and secondary analyses described in Section \ref{subsec: primary analysis} to \ref{subsec: secondary analysis II dose-response relationship} on $462$ matched pairs of $2$ non-rural counties and $749$ matched pairs of $2$ rural counties. P-values when testing the primary analysis hypothesis $H_{0, \text{primary}}$ concerning the death toll and secondary analysis hypothesis $H_{0, \text{secondary}}$ concerning the case number are $0.004$ and less than $10^{-5}$, respectively, in the non-rural subgroup, and $0.008$ and $0.009$, respectively, in the rural subgroup. We also allowed a differential dose-response relationship between social distancing and case numbers in rural and non-rural counties and constructed confidence sets for $(\tau, \beta)$ separately; see Figure \ref{fig: real data metro and rural contour and dose-response curves}. We further repeated the subgroup analyses under different specifications of the cumulative dose and the results were similar; see Supplementary Material G.3.4 for details.

A comparison of confidence sets for the non-rural counties (top left panel of Figure \ref{fig: real data metro and rural contour and dose-response curves}) and rural counties (bottom left panel of Figure \ref{fig: real data metro and rural contour and dose-response curves}) revealed an intriguing pattern: while the level-$0.1$ confidence set of the activation threshold $\tau$ is centered around the range of $0.4-0.6$ for the rural counties, it contains $0$ for the non-rural counties; moreover, the level-$0.1$ confidence set of rural counties covers a much larger range of $\beta$ values compared to that of the non-rural counties. Together, these results suggest that the activation dose required to trigger exponential growth in case numbers in rural counties seemed much larger than that in non-rural counties; however, once exponential growth in case numbers was incurred, the growth seemed more rapid in rural counties. 

\subsection{Dose-response relationship under local interference}
\label{subsec: real data dose-response under interference}
We next applied the methodology developed in Section \ref{sec: interference} and \ref{subsec: incorporate interference longitudinal} to obtain corrected confidence sets of $(\tau, \beta)$ under local interference. To this end, we collect $2 \times 1,211$ copies of reference dose trajectory $\mathbf{z}_{t_0:T}^\ast$ in $\vec{\mathbf{z}}_{t_0:T}^\ast$ and the cumulative doses of all study units during the treatment period in $\vec{\mathbf{z}}_{\text{cumu}} = (\text{CD}(\mathbf{z}_{11; t_0: T}; \mathbf{z}^\ast_{t_0: T}, \mathcal{W}_{\text{lag}}), \cdots, \text{CD}(\mathbf{z}_{I2; t_0: T}; \mathbf{z}^\ast_{t_0: T}, \mathcal{W}_{\text{lag}}))$. We consider relaxing the dose-response relationship by incorporating local interference as follows:
\begin{equation*}
\begin{split}
    H_{0, \text{kink}, \text{interference}}: ~~&Y_{ij, \text{case}, \text{agg}}(\vec{\mathbf{z}}_{t_0: T}) - Y_{ij, \text{case}, \text{agg}}(\vec{\mathbf{z}}_{t_0:T}^\ast) = 0,~\forall \vec{\mathbf{z}}_{t_0: T} ~\text{such that}~ \text{CD}(\mathbf{z}_{ij; t_0: T}; \mathbf{z}^\ast_{t_0: T}, \mathcal{W}_{\text{lag}}) \leq \tau, \\ \text{and}~~&\log\{Y_{ij,\text{case}, \text{agg}}(\vec{\mathbf{z}}_{t_0: T})\} - \log\{Y_{ij, \text{case}, \text{agg}}(\vec{\mathbf{z}}^\ast_{t_0:T})\}\\
    = &\beta\cdot\{\text{CD}(\mathbf{z}_{ij; t_0: T}; \mathbf{z}^\ast_{t_0: T}, \mathcal{W}_{\text{lag}}) - \tau\} \cdot\bigg\{1 + \underbrace{\frac{1}{1 + \exp\{-k( \|\mathbf{G}_{ij, \bigcdot}\|^{-1}_0\cdot\langle\vec{\mathbf{z}}_{\text{cumu}}, \mathbf{G}_{ij,\bigcdot}\rangle - s)\}}}_{\text{Spillover Effect Factor} ~C}\bigg\},\\
    &\forall\vec{\mathbf{z}}_{t_0: T}~\text{such that}~\text{CD}(\mathbf{z}_{ij; t_0: T}; \mathbf{z}^\ast_{t_0: T}, \mathcal{W}_{\text{lag}}) > \tau,~~\forall i = 1, \cdots, I = 1211, j = 1, 2.
\end{split}
\end{equation*}
According to this null hypothesis, there is no direct effect if county $ij$'s cumulative dose is below some threshold $\tau$; hence, there is no spillover effect in this case by Principle III described in Section \ref{sec: interference}. Once county $ij$'s cumulative dose is above the threshold, this triggers exponential growth captured by the dose-response direct effect plus a spillover effect. The magnitude of the spillover effect is equal to the direct effect multiplied by a spillover effect factor $C$. This spillover effect factor depends on the average cumulative dose of $ij$'s neighbors but is always upper bounded by $1$ so that the spillover effect is no larger than the direct effect (see Section \ref{subsec: dose-response under interference}). In rare cases when a county has no neighbor, $C$ is defined to be $0$ so that there is no spillover effect. We used the county adjacency file provided by the United States Census Bureau (\citealp{US_census_adj}) as our adjacency matrix $\mathbf{G}$. 

The interference parameters $(k, s)$ in the above model are easy to interpret and specify. For instance, $(k, s) = (5.0, 1.0)$ corresponds to a small spillover effect of approximately $1\%$ of the direct effect when neighbors' average cumulative dose is $0.10$ (corresponding to an average $40\%$ reduction in social mobility compared to the pre-pandemic level during the treatment period) and approximately $8\%$ of the direct effect when neighbors' average cumulative dose is $0.50$ (corresponding to social mobility remaining the same as the pre-pandemic level during the treatment period). In this way, the interference parameters $(k, s)$ carry concrete meanings and can be easily tuned and communicated to the audience.

\begin{figure}[h]
    \centering
    \subfloat{\includegraphics[width = 0.49\columnwidth]{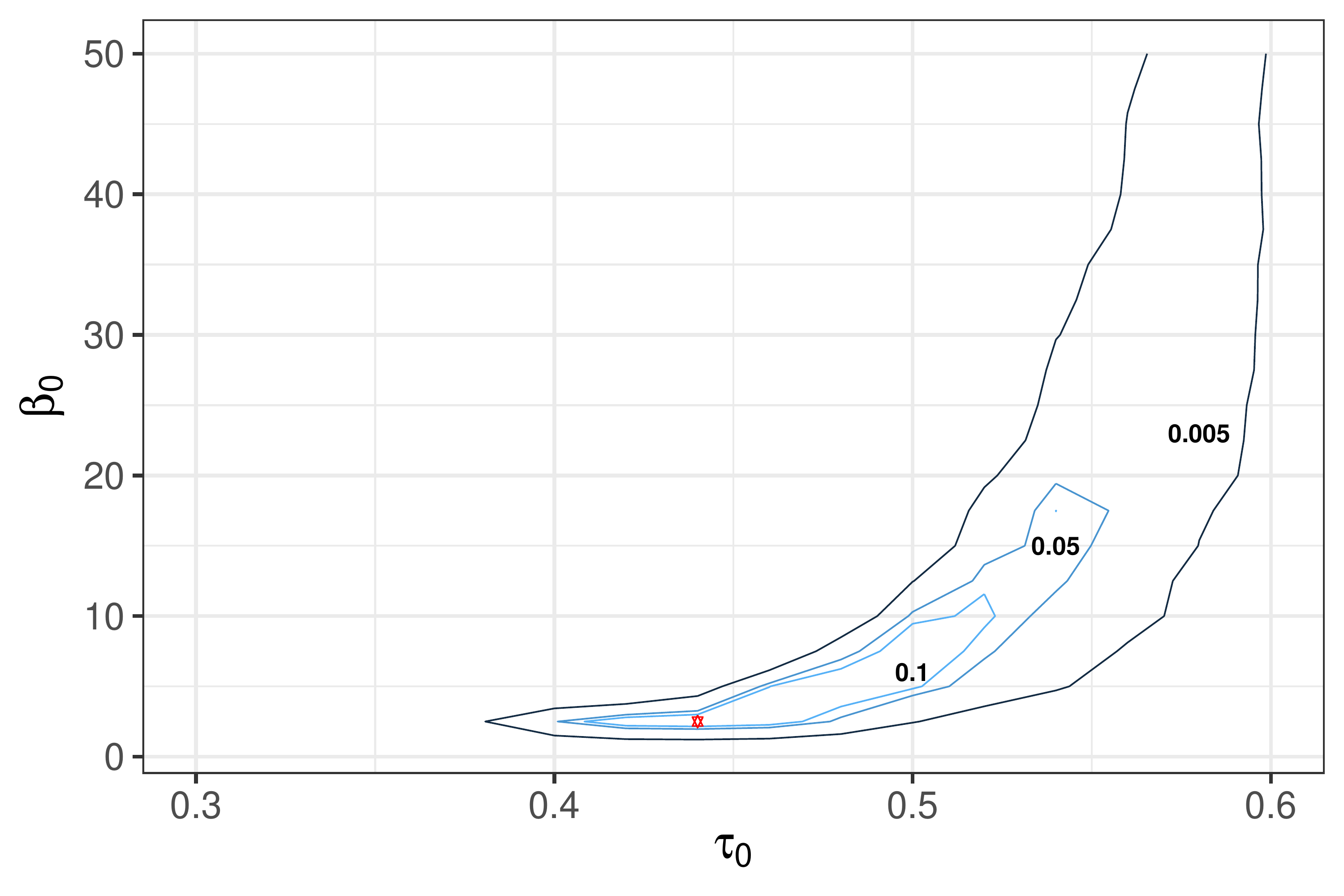}}
     \subfloat{\includegraphics[width = 0.49\columnwidth]{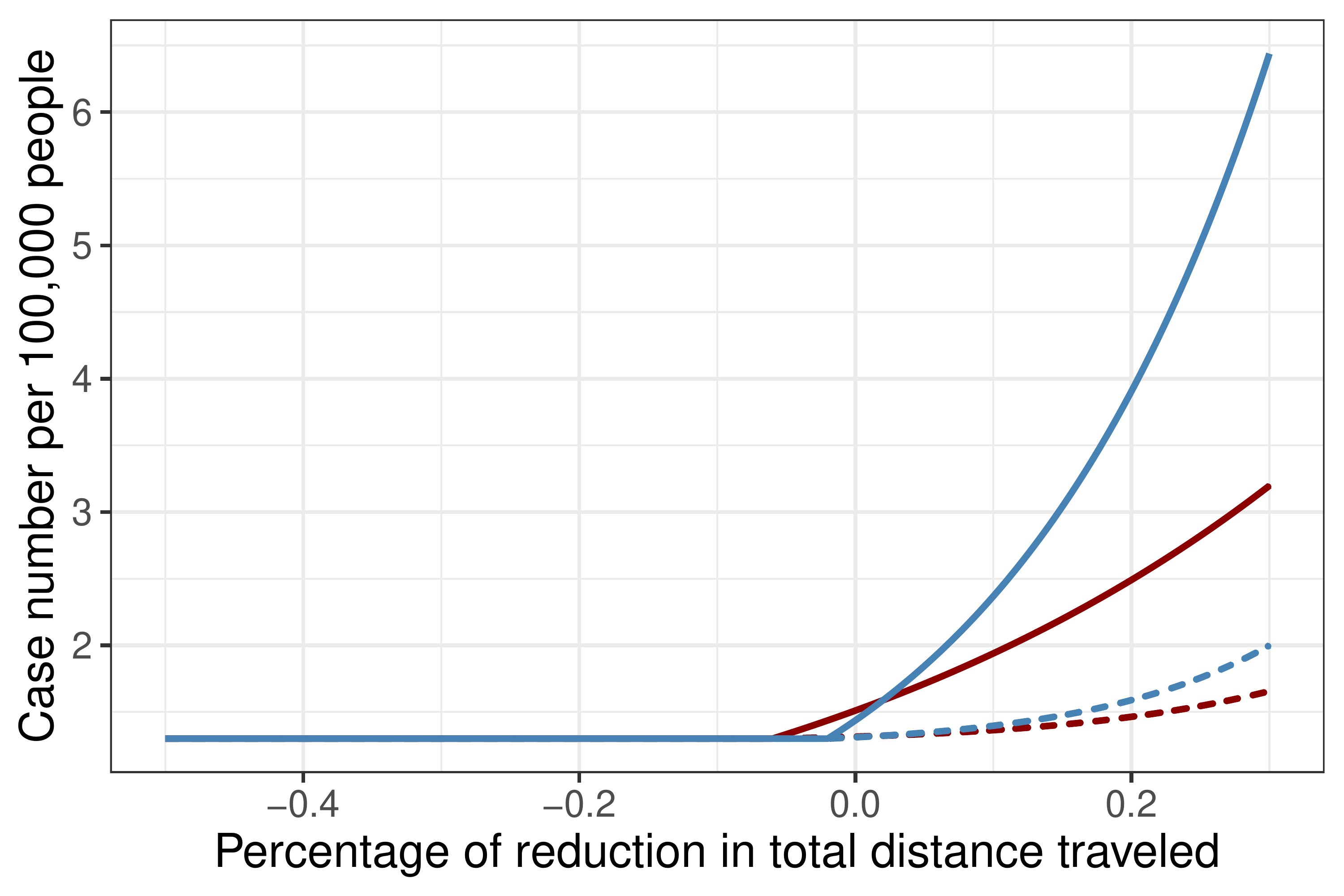}}
    \caption{\small Left panel: contour plot of p-values when testing $H_{0, \text{kink}, \text{interference}}$ against $\tau = \tau_0$ and $\beta = \beta_0$ under interference parameters $(k, s) = (5, 1)$. Maximum p-value is obtained at $(\tau_0, \beta_0) = (0.44, 2.5)$ (red marker). Three isopleths ($0.1$, $0.05$, and $0.005$) are plotted. Right panel: dose-response direct effect (solid lines) and the associated spillover effects (dotted lines) for selected $(\tau_0, \beta_0)$ in the $0.05$ confidence set with baseline $Y_{ij, \text{case}, \text{agg}}(\mathbf{z}^\ast_{t_0: T})$ equal to $1$ per $100,000$. Two red lines correspond to $(\tau_0, \beta_0) = (0.44, 2.5)$ and blue lines $(\tau_0, \beta_0) = (0.48, 5.0)$.}
    \label{fig: real data interference}
\end{figure}

The left panel of Figure \ref{fig: real data interference} plots the level-$0.1$, $0.05$, and $0.005$ confidence sets of $(\tau, \beta)$ when the interference parameters $(k, s) = (5, 1)$. The right panel of Figure \ref{fig: real data interference} further illustrates the inferred dose-response direct effects (solid lines) and the associated spillover effects (dotted lines) under $(k, s) = (5, 1)$ for $(\tau, \beta) = (0.44, 2.5)$ (red lines) and $(0.48, 5.0)$ (blue lines).

The level-$0.05$ confidence set of the dose-response direct effect contains similar $\tau$ values but considerably smaller $\beta$ values compared to assuming no interference and modeling the total effect using a dose-response kink model (see the left panel of Figure \ref{fig: real data secondary analysis contour and dose-response curves}). This makes intuitive sense as the total effect has now been decomposed into the dose-response direct effect and a spillover effect due to neighboring counties. 

\section{Discussion}
\label{sec: discussion}
We studied in detail the effect of social distancing during the early phased reopening in the United States on COVID-19 related death toll and case numbers using our compiled county-level data. To address the statistical challenge brought by a time-dependent, continuous treatment dose trajectory, we developed a design-based framework based on nonbipartite matching to embed observational data with time-dependent, continuous treatment dose trajectory into a randomized controlled experiment. This embedding induces a randomization scheme that we then used to conduct randomization-based, model-free statistical inference for causal relationships, including testing a causal null hypothesis, a structured dose-response relationship and a causal null hypothesis under local interference modeling.

Upon applying the proposed design and testing procedures to the mobility and COVID-19 data, we found very strong evidence against the causal null hypothesis and supportive of a causal effect of social distancing during the early phases of reopening on subsequent COVID-19-related death and case numbers. Our finding complements many recent studies based on standard epidemiological models (\citealp{dickens2020institutional, koo2020interventions}) and structural equation modeling (\citealp{chernozhukov2021causal, bonvini2021causal}) from a unique perspective, and once again confirms the important role of social distancing (as captured by a reduction in mobility in this article) in combating the novel coronavirus. Our transparent comparison of two groups of similar counties makes our findings digestible and easy to communicate to the general public.

In a dose-response analysis, we found that the confidence set of the dose needed to activate exponential growth was tightly centered and its magnitude suggested that once the total distance traveled returned to or even superseded the pre-coronavirus level, it would have a devastating effect on the COVID-19 case numbers by contributing to exponential growth. Moreover, in a subgroup analysis where we allowed a differential dose-response relationship, we found that more stringent social distancing would be needed to avoid devastating exponential growth for non-rural counties; however, once the exponential growth was incurred, the growth appeared more rapid in rural counties. This striking difference in dose-response relationship between rural and non-rural communities agrees with experts' assessment of the transmission dynamics. Given its clinical features, the rate of virus reproduction is likely higher in large, urban areas due to more reproductive opportunities afforded by denser populations (\citealp{souch2020commentary}) and this may explain the absence of an ``activation dose" in non-rural counties (see top left panel of Figure \ref{fig: real data metro and rural contour and dose-response curves}). On the other hand, although rural residents have less social interaction compared to non-rural counterparts, they often have more underlying medical conditions and are more likely to present for treatment at more advanced stages of disease (\citealp{callaghan2021rural}), which may partly explain why rural communities seemed to incur more drastic exponential growth in case numbers once the activation dose was exceeded (see bottom left panel of Figure \ref{fig: real data metro and rural contour and dose-response curves}).

The design-based approach and analysis proposed in this article has its limitations. First, we used social mobility data as a proxy measure for social distancing. It would be interesting to look at other aspects of social distancing, e.g., closure of borders, reduction in aviation travel, etc, in future works. Second, in order to permute two treatment dose trajectories in a longitudinal setting, one necessarily needs to match on observed outcomes during the treatment period and compare outcomes after the treatment period; therefore, in a longitudinal setting, the method is suited only for applications where the effect of a time-varying treatment is not immediate, e.g., effect of precautionary measures on the death toll. Third, when the sample size is limited, the interference parameters are treated as sensitivity parameters that researchers vary, rather than population parameters for which researchers construct confidence sets. The proposed method also has its unique strengths: it embeds the noisy observational data into an approximate randomized controlled trial and has a clear ``reasoned basis" (\citealp{Fisher1935design}) when testing the causal null hypothesis, and researchers can always conduct a sensitivity analysis to investigate how causal conclusions would change when the randomization assumption is relaxed. The method developed in this article can be readily applied to many practical problems where there is a continuous exposure and the scientific interest lies in testing a dose-response relationship. Understanding a dose-response relationship is central to many scientific disciplines like public health (\citealp{gorell1999smoking, farrelly2005evidence}), pharmacology (\citealp{tallarida2012dose}), and toxicology (\citealp{calabrese2003hormesis}), among many others.

\begin{appendix}
\section{ Map of $1,211$ better and $1,211$ worse social distancing counties}\label{appA}

\begin{figure}
    \centering
    \includegraphics[width=\textwidth]{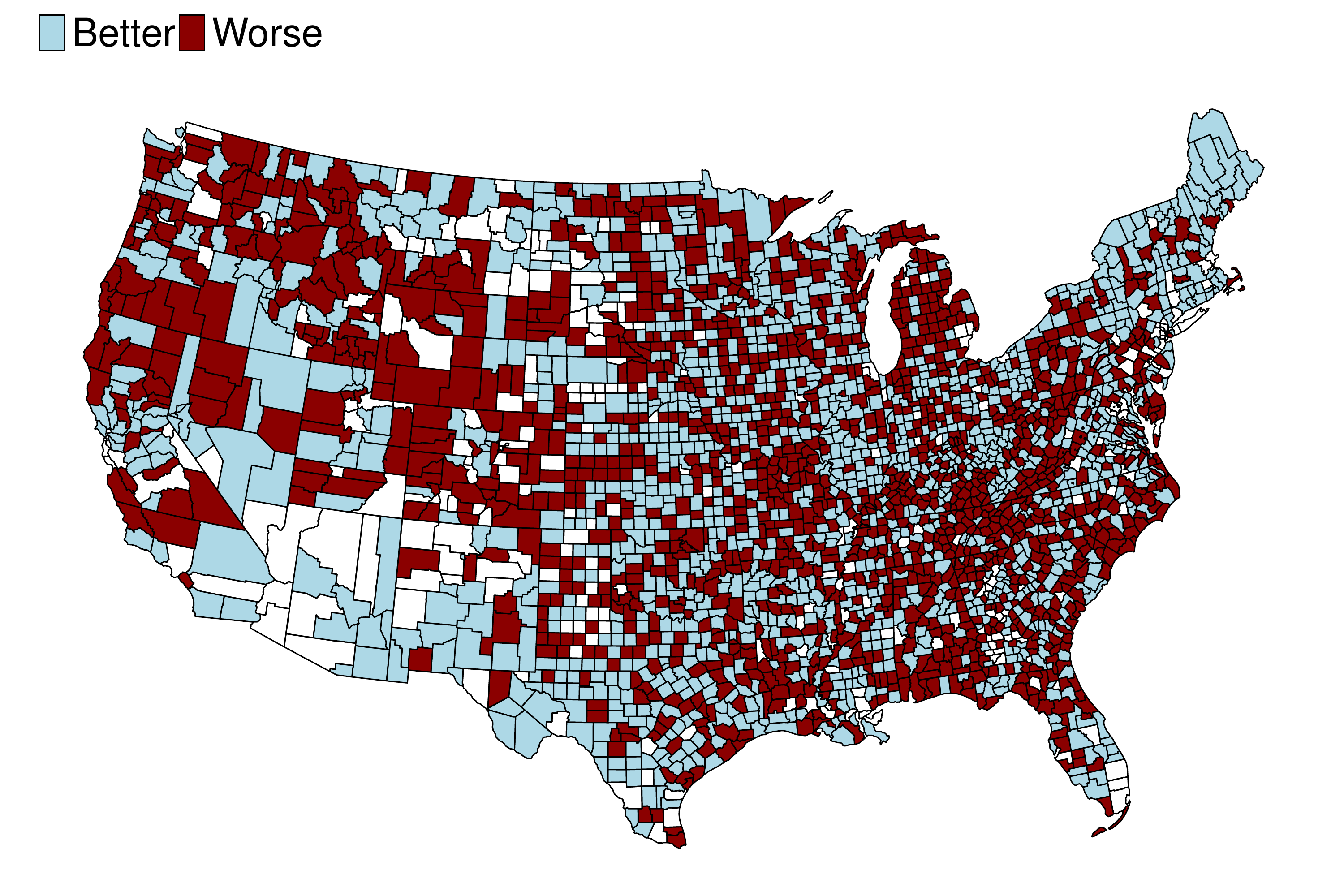}
    \caption{Map of $1,211$ better social distancing (light blue) and $1,211$ worse social distancing counties (red) in the matched analysis. Unmatched counties are in white.}
    \label{fig: map}
\end{figure}


\section{Balance table after statistical matching}\label{appB}
\begin{table}[h]
\centering
\label{tbl: nbpmatch balance overall}
\resizebox{\textwidth}{!}{
\begin{tabular}{lccccc}
  \hline
 &\multirow{3}{*}{\begin{tabular}{c}Better Social \\ Distancing Counties\\ (n = 1,211)\end{tabular}}  
 & \multirow{3}{*}{\begin{tabular}{c}Worse Social\\ Distancing Counties \\ (n = 1,211)\end{tabular}} 
 & \multirow{3}{*}{\begin{tabular}{c}Standardized \\Difference\end{tabular}}  \\  \\ \\
  \hline \\ \vspace{-0.9 mm}
  \textbf{Time-Independent Covariates} \\ \\ \vspace{-0.9 mm}
  \hspace{0.3 cm} female (fr) & 0.50 & 0.50 & 0.06  \\ 
  \hspace{0.3 cm} above 65 (fr) & 0.20 & 0.19 & -0.05 \\ 
  \hspace{0.3 cm} black (fr) & 0.07 & 0.07 & 0.00    \\ 
  \hspace{0.3 cm} hispanic (fr) & 0.09 & 0.08 & -0.03  \\ 
  \hspace{0.3 cm} driving alone to work (fr) & 0.80 & 0.81 & 0.12    \\ 
  \hspace{0.3 cm} smoking (fr) & 0.17 & 0.18 & 0.15 \\ 
  \hspace{0.3 cm} flu vaccination (fr) & 0.42 & 0.42 & -0.01   \\ 
  \hspace{0.3 cm} some college (fr) & 0.59 & 0.58 & -0.09   \\ 
  \hspace{0.3 cm} membership association (per 10,000 people) & 12.29 & 12.01 & -0.05 \\ 
  \hspace{0.3 cm} rural (0/1) & 0.62 & 0.62 & 0.00 \\ 
  \hspace{0.3 cm} below poverty (fr) & 0.14 & 0.15 & 0.14  \\ 
  \hspace{0.3 cm} population density (residents per $\text{mi}^2$) & 173 & 130 & -0.08   \\ 
  \hspace{0.3 cm} population & 92,310 & 79,423 & -0.06   \\  \\  \vspace{-0.9 mm}
  
  \textbf{Time-Varying Covariates (per 100,000 people)} \\ \\ \vspace{-0.9 mm}
  \hspace{0.3 cm} Cases during Apr 20th - Apr 26th  & 27.97 & 25.30 & -0.02     \\ 
  \hspace{0.3 cm} Cases during Apr 27th - May 3rd & 29.74 & 24.23 & -0.08   \\ 
  \hspace{0.3 cm} Cases during May 4th - May 10th  & 29.17 & 25.25 & -0.05   \\ 
  
  \hspace{0.3 cm} Cases during May 11th - May 17th  & 26.21 & 25.49 & -0.01   \\ 
  \hspace{0.3 cm} Cases during May 18th - May 24th  & 29.13 & 28.70 & -0.01    \\ 
  \hspace{0.3 cm} Cases during May 25th - June 1st  & 30.95 & 25.62 & -0.07    \\ 
  \hspace{0.3 cm} Cases during June 2rd - June 8th  & 29.75 & 28.78 & -0.01    \\ 
  
  \hspace{0.3 cm} Cases during June 9th - June 15th   & 28.40 & 31.31 & 0.04    \\ 
  \hspace{0.3 cm} Cases during June 16th - June 22th  & 34.02 & 40.97 & 0.09     \\ 
  \hspace{0.3 cm} Cases during June 23th - June 29th  & 45.51 & 51.94 & 0.09    \\ 
  
  \hspace{0.3 cm} Deaths during Apr 20th - Apr 26th  & 1.35 & 0.92 & -0.12     \\ 
  \hspace{0.3 cm} Deaths during Apr 27th - May 3rd  & 1.20 & 0.95 & -0.08     \\ 
  \hspace{0.3 cm} Deaths during May 4th - May 10th  & 1.35 & 1.00 & -0.09     \\ 
  
  \hspace{0.3 cm} Deaths during May 11th - May 17th  & 1.00 & 0.98 & -0.01    \\ 
  \hspace{0.3 cm} Deaths during May 18th - May 24th  & 0.95 & 0.91 & -0.01     \\ 
  \hspace{0.3 cm} Deaths during May 25th - June 1st  & 1.06 & 0.84 & -0.07     \\ 
  \hspace{0.3 cm} Deaths during June 2rd - June 8th  & 0.85 & 0.70 & -0.06    \\ 
  
  \hspace{0.3 cm} Deaths during June 9th - June 15th  & 0.67 & 0.66 & -0.00     \\ 
  \hspace{0.3 cm} Deaths during June 16th - June 22th & 0.62 & 0.64 & 0.01     \\ 
  \hspace{0.3 cm} Deaths during June 23th - June 29th & 0.85 & 0.68 & -0.05     \\ 
   \hline
\end{tabular}}
\end{table}

\end{appendix}

\begin{supplement}
\stitle{Pilot study, technical details, and further details on the case study}
\sdescription{Supplementary Material A provides details on the pilot study described in Section 1.1 in the main article. Supplementary Material B motivates the Kolmogorov-Smirnov-type test statistic considered in the main article. Supplementary Material C discusses how to construct a confidence set for nuisance parameters $(\tau, \beta)$ in a dose-response kink model based on a variant of rank sum test. Supplementary Material D illustrates how to test a sequence of dose-response relationship ordered according to their model complexity. Supplementary Material E derives the treatment dose trajectory assignment probability in each matched pair. Supplementary Material F provides details on generalizing the dose-response relationship to an aggregate outcome. Supplementary Material G provides further details on the case study, including maps of the $1,211$ better and worse social distancing counties in the matched samples, the balance table, a closer examination of the distributions of some important variables after matching, separate analyses of rural and non-rural counties, and numerous sensitivity analyses. Supplementary Material H assesses Assumption \ref{assump: potential outcome same under CD} using a standard epidemiological model.}
\end{supplement}
\begin{supplement}
\stitle{code and data.zip}
\sdescription{Data and \textsf{R} code implementing the statistical matching and randomization inference.}
\end{supplement}

\bibliographystyle{imsart-nameyear} 
\bibliography{paper-ref}       


\end{document}